\documentclass[authoryear,12pt,a4paper]{elsarticle}
\usepackage{multirow}
\usepackage{epsfig, setspace, pmgraph}
\usepackage{wrapfig}
\usepackage[lofdepth,lotdepth]{subfig}
\usepackage{graphicx}
\usepackage{afterpage}
\usepackage{amsmath, amsfonts, amssymb}
\usepackage{dsfont,courier}
\usepackage{verbatim}
\usepackage{upgreek}
\usepackage{longtable}
\usepackage{color,soul}
\usepackage[modulo]{lineno}
\usepackage{caption}
\usepackage{booktabs}
\usepackage[letterpaper,margin=25 mm]{geometry}%
\usepackage{pdfpages}
\usepackage{calc}
\usepackage{enumitem}
\modulolinenumbers[2]

\newcommand{\tab}[1]{%
{\makebox[.12\linewidth][l]{#1}\ignorespaces}%
}

\journal{Journal of Materials Processing Technology}
\begin{document}

\begin{frontmatter}

\title{Analysis and modelling of a rotary forming process for cast aluminum alloy A356}

\author[manx]{M.J. Roy\corref{cor1}}
\ead{matthew.roy@manchester.ac.uk}
\author[ubc]{D.M. Maijer}
\ead{daan.maijer@ubc.ca}

\address[manx]{School of Mechanical, Aerospace and Civil Engineering, The University of Manchester, Manchester, UK, M13 9PL}

\address[ubc]{Dept. of Materials Engineering, The University of British Columbia, Vancouver, BC, Canada V6T 1Z4}

\cortext[cor1]{Corresponding author, Tel. +44 (0)161 275 4316}

\begin{abstract}
Spinning of a common aluminum automotive casting alloy A356 (Al-7Si-0.3Mg) at elevated temperatures has been investigated experimentally with a novel industrial-scale apparatus. This has permitted the implementation of a fully coupled thermomechanical finite element model aimed at quantifying the processing history (stress, strain, strain-rate and temperature) and predicting the final geometry. The geometric predictions of this model have been compared directly to the geometry of the workpieces obtained experimentally. This study is novel in regards to both the size and shape of the component as well as the constitutive material representation employed.  The model predictions are in reasonable agreement with experimental results for small deformations, but errors increase for large deformation conditions. The model has also enabled the characterization of the mechanical state which leads to a common spinning defect. Suggestions for improving the accuracy and robustness of the model to provide a predictive tool for industry are discussed.
\end{abstract}


\end{frontmatter}

\clearpage

\section*{Nomenclature}

\begin{description}[noitemsep]
\item[\tab{$P$}] Radial penetration of roller into workpiece employed in model \hfill (mm)%
\item[\tab{$L$}] Element edge length \hfill (mm)%
\item[\tab{$k_c$}] Thermal conductivity \hfill ($\text{W} \text{ m}^{-1}\,^{\circ}\text{C}^{-1}$)%
\item[\tab{$C_p$}] Specific heat capacity\hfill ($\text{J} \text{ kg}^{-1}\,^{\circ}\text{C}^{-1}$)%
\item[\tab{$\rho$}] Density\hfill ($\text{kg} \text{ m}^{-3}$)%
\item[\tab{$\alpha$}] Coefficient of thermal expansion \hfill ($^{\circ}$C$^{-1}$)%
\item[\tab{$\alpha'$}] Total thermal strain %
\item[\tab{$T_0$}] Reference temperature for thermal expansion\hfill ($^{\circ}$C)%
\item[\tab{$T_m$}] Mandrel temperature employed in model\hfill ($^{\circ}$C)%
\item[\tab{$T_b$}] Workpiece temperature employed in model\hfill ($^{\circ}$C)%
\item[\tab{$u_o,\,u_n,\,u_f$}] Axial/radial roller nose positions employed in model \hfill (mm)%
\item[\tab{$t_p$}] Simulated process time\hfill (s)%
\item[\tab{$\beta$}] Taylor-Quinney factor%
\item[\tab{$q_{c,\,f}$}] Heat fluxes applied to model workpiece cooling \hfill ($\text{W} \text{ m}^{-2}$)%
\item[\tab{$\sigma_1$}] Maximum principal stress\hfill (MPa)%
\item[\tab{$\sigma_{\text{VM}}$}] von Mises equivalent stress\hfill (MPa)%
\item[\tab{$F_c$}] Clamping force employed in model\hfill (kN)%
\end{description}

\clearpage
\section{Introduction}\label{sec:intro}
Often manufacturing technologies can be combined to leverage specific advantages inherent to each. Axisymmetric and complex shaped components can be formed through near-net shape casting processes alone. However, designers must account for the microstructural inhomogeneities and potential defects that are inherent to cast components. Standard, monolithic forging of castings, such as through drawing or stamping, will ameliorate mechanical property variations, however these process have prohibitive tooling and production costs, which act as significant barriers to their use. As such, adopting incremental rotary forming practices, namely flow forming, shear forming and / or spinning, are ideal complimentary processing techniques for castings with axisymmetric geometries. Incremental forming is generally a more economical manufacturing methodology over forming process that are applied to entire components, such as standard forging, because forming loads and friction are minimized and simpler, lower cost tooling is needed.

Even in light of these advantages, industry has been slow to adopt incremental rotary forming processes owed to often costly trial-and-error approach to process development. Particularly when forming at elevated temperatures, there are a myriad of process parameters: forming speeds, tooling geometry, material properties and history all play a role in deciding the final component geometry and in-service performance. Robust process models for industry are required to enable virtual fabrication simulation, and thereby decrease process development time, if the benefits of incremental forming are to be realized with any impact.

Rotary forming processes consist of spinning an axisymmetric workpiece through a set of impinging tools to locally plasticize the material. Usually these tools to generate a plastic zone between a rigid mandrel supporting the inner workpiece diameter and a roller in contact with the outer diameter of the workpiece. This zone moves with the roller or tool, with typical applications having the material repeatedly encountering the same plastic zone to achieve bulk deformation. The term `rotary forming' has been used to describe this class of incremental forming processes in the literature. Listed in order of degree of deformation delivered to the work piece, flow forming, tube spinning, shear forming and spinning are all covered by this description. For additional, more nuanced process descriptions the reader is directed to the literature review conducted by \cite{Wong.03} focussing on all of the aforementioned processes and provides detailed descriptions of each of the processes beyond deformation levels. More recently, \cite{Music.10} has conducted an extensive review focussing on the mechanics of spinning processes.

The application of rotary forming techniques to Al-Si-Mg cast aluminum alloys has been studied previously by \cite{Mori.09} and \cite{Cheng.10}. Both authors conducted spinning trials at elevated temperatures. Both studies showed an amelioration in ductility and slight changes in the yield strength post processing and tempering. Cheng et al. documented that there was a slight improvement in corrosion resistance beyond improvement of mechanical properties. Mori et al. found that porosity was eliminated from A357 alloy blanks machined from larger castings and formed at temperatures between 350 and 400$^{\circ}$C. Mori et al. also reported that mechanical properties improved during spinning at lower temperatures, however process-induced flaws appeared in the form of surface cracks. Their formation was investigated using a model developed with the commercial explicit finite element analysis, LS-DYNA. The sole correlation made with predictions from this isothermal process model was that cracking occurred in regions with high levels of plastic strain.

To the knowledge of the authors, there has been no previous attempt to produce a fully coupled thermomechanical model of a spinning operation which incorporates rate-dependent material properties. For aluminum alloys, the effects of strain rate and temperature have significant effects on the response of the material, which has not been completely captured in previous studies. For incremental forming operations, small regions of the workpiece will undergo high strain rate deformation and accompanying temperature increases due to plastic deformation, with the surrounding material essentially unaffected. While there are multiple modelling efforts of this process documented, most have employed isothermal models and flow stress descriptions found by quasi-static testing. While \cite{Klocke.04} used different FEA packages to capture the sequential thermal and mechanical effects on spinning of stainless steel, this approach does not capture the heat generated due to plastic deformation. Furthermore, Klocke and Wehrmeister did not detail the degree of coupling between the thermal and mechanical aspects of their model; for example, detailing at what frequency the thermal and mechanical solutions were updated. Most recently, \cite{Mohebbi.10} employed ABAQUS Explicit to generate a flow forming model for an annealed wrought aluminum alloy (AA6063). However, this model was only validated by qualitative comparisons to experimental results where forming loads were compared to analytical expressions derived for a different material. 

\cite{Roy.14} have previously employed a novel rotary forming apparatus to perform a series of experiments to ascertain the effects of both forming and subsequent heat treatment on the microstructure of A356. This study focussed on determining the effects of forming on precipitation and secondary phases and overall implications for further heat treatment. The current paper will document how these experiments have formed the basis of a modelling effort aimed at determining the evolution of strain, strain-rate and temperature during rotary forming at elevated temperatures. \cite{Roy.12} have previously characterised the constitutive behaviour of as-cast A356. In this latter work, a number of constitutive expressions to describe the material at various temperatures, strain rates and deformation levels were developed with varying degrees of efficacy of describing the actual mechanical response. This material description has been leveraged to develop the forming model described presently. The results of the model are compared to rotary forming experiments which cover a range of forming conditions and can be best described as spinning-on-air. In some cases, the deformation imposed was found to be more than that expected by standard spinning and so as a point of convenience, the term spinning and rotary forming are used interchangeably. As there is a relative paucity of studies quantifying the degree to which standard FEA-based models of spinning compare to experimental results, the geometric predictions of the model are compared directly to the geometry of the workpieces obtained experimentally. This comparison is followed by discussion regarding proposed improvements targeting the model's efficacy and accuracy from the standpoint of industrial users as a predictive tool.

\section{Experimental forming methodology}\label{sec:Exp}
\begin{figure}
\centering
\subfloat[\label{fig:topViewProcess}]{\includegraphics[width=0.3922\linewidth]{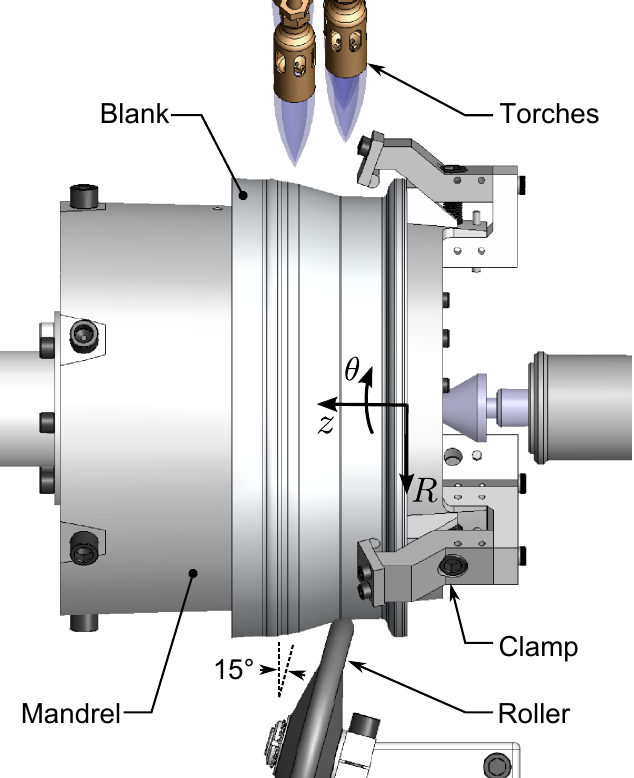}}\hfill
\subfloat[\label{fig:formSpeedsFeeds}]{\includegraphics[width=0.5\linewidth]{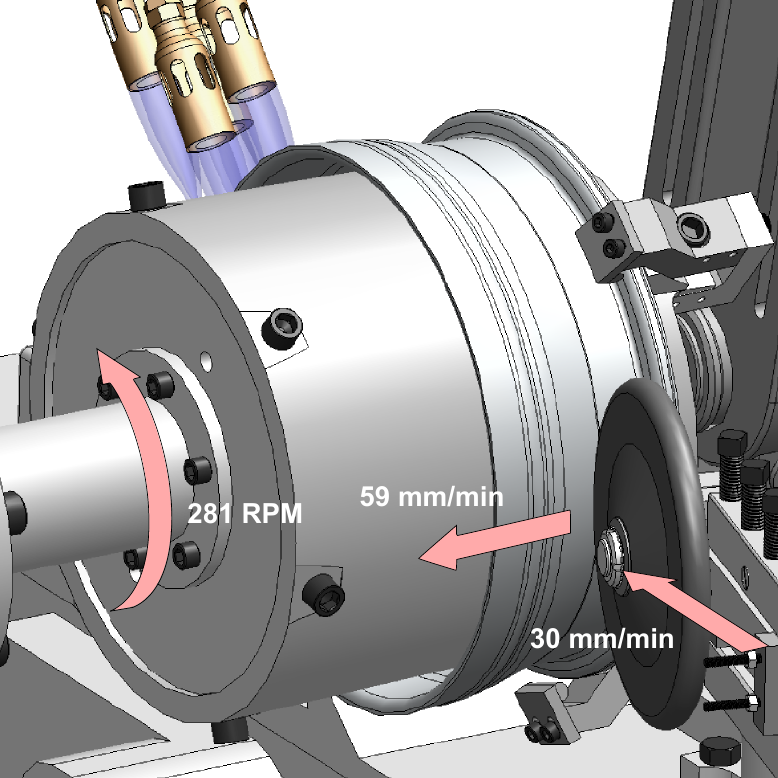}}
\caption{A top-down view of specific apparatus components in (a) and a parametric depiction with forming directions and speeds in (b).}
\label{fig:ProcDesc}
\end{figure}

An experimental forming apparatus was developed to perform spinning experiments at elevated temperatures with workpieces representing the same geometry as an automotive wheel. The apparatus consists of a belt-driven lathe driven by a 30 HP motor, a bank of propane torches to heat the workpiece with approximately 82 kW of total heat output and tooling in the form of a mandrel and roller assembly. The roller, made from AISI-8620 steel, had a diameter of 120 mm, a nose radius of 10 mm and was held rigidly 15$^{\circ}$ off of the radial axis of the workpiece (Fig. \ref{fig:ProcDesc}).

\begin{figure}[]
\centering
\includegraphics[width=0.7464\linewidth]{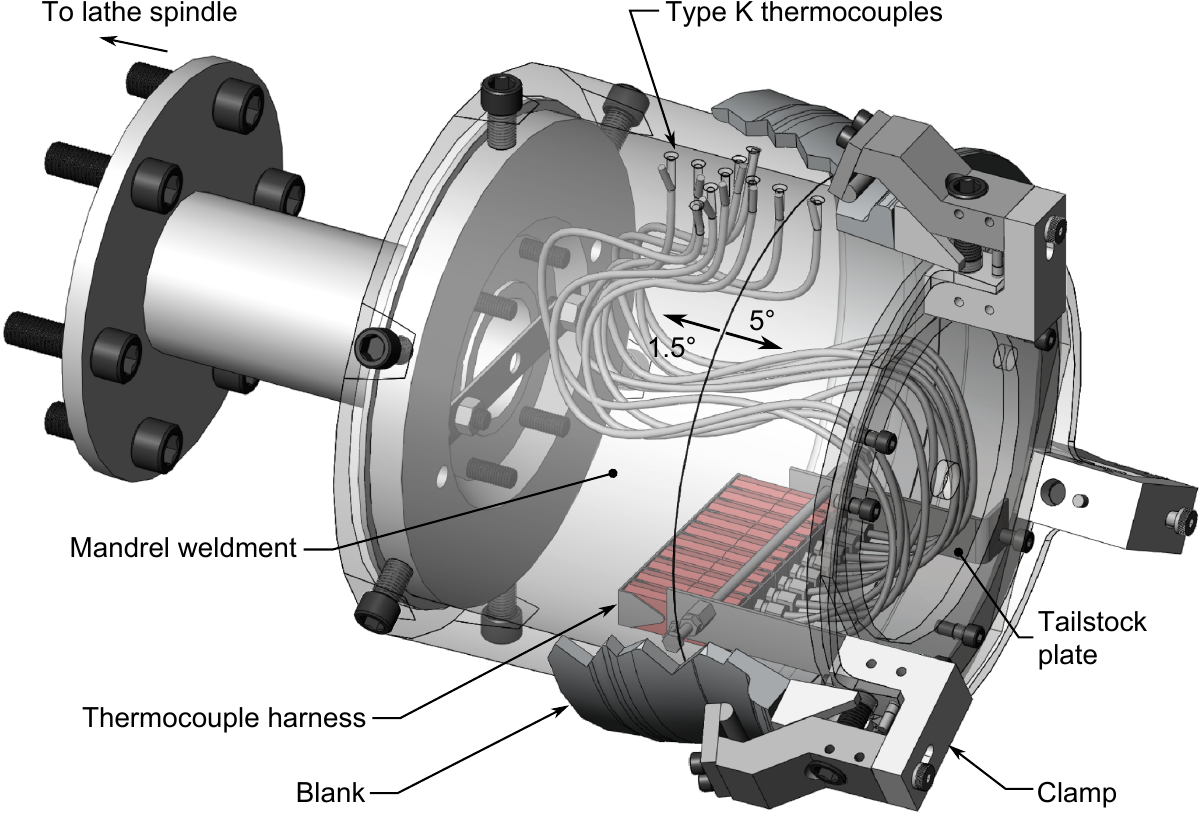}
\caption{Depiction of the rotary tooling assembly.}
\label{fig:Mandrel}
\end{figure}

The mandrel was constructed from AISI-4130 and designed to account for the workpiece's thermal expansion and hold the workpiece rigidly during forming, matching the workpiece's ID (Fig. \ref{fig:Mandrel}). The design required that the workpiece was to be heated to forming temperatures from ambient conditions in situ, while held in place by manually actuated, captive clamps. The mandrel was tapered to ease the release of the formed workpieces, and account for thermal expansion of the workpiece. The mandrel was bolted directly to the spindle of the lathe on the non-forming end, and supported by a live center on the tailstock end, such that the outer diameter of the mandrel had 0.5 mm circular runout. The mandrel was furthermore instrumented with a series of type-K thermocouples (TCs) that were read via a wireless data acquisition system fixed to the quill of the lathe. In conjunction with TCs peened to a single commissioning workpiece, these transducers were employed to capture the thermal profile of the forming process.

As-cast (AC) A356 (Al-7wt\%Si-0.3wt\%Mg) wheels, produced via the low-pressure die casting process, were provided by a North American wheel manufacturer. Workpieces or blanks suitable for spinning were machined from these wheels by removing the spokes and hub from the wheel, but preserving the main annulus. The resulting workpiece measured approximately 140 mm axially, had a minimum internal diameter of 330 mm and an approximate wall thickness of 10 mm.

Prior to performing forming experiments, the mandrel was preheated to approximately 150$^{\circ}$C with the heating system, and then sprayed with a colloidal graphite coating to act as a lubricant. Workpieces were preheated to the same temperature in a box furnace, and the inner diameter was sprayed with refractory-type coating (Foseco DYCOTE 32). This coating was applied to assist in workpiece removal as well as to suppress heat transfer between the workpiece and the mandrel, diminishing workpiece heating times. The outer diameter of the workpiece was coated with the same colloidal graphite as the mandrel outer diameter. The workpiece and tooling were then allowed to cool to ambient conditions prior to commencing forming trials.

\subsection{Thermal profile}
\begin{figure}
\centering
\subfloat[Thermocouple locations\label{fig:TCloc}]{\includegraphics[width=0.4516\linewidth]{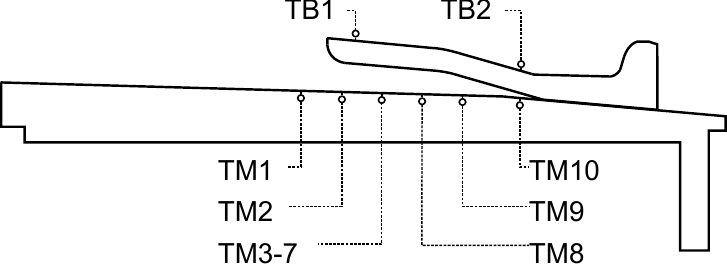}}\\
\subfloat[Mandrel temperatures \label{fig:TCMand}]{\includegraphics[width=0.5\linewidth]{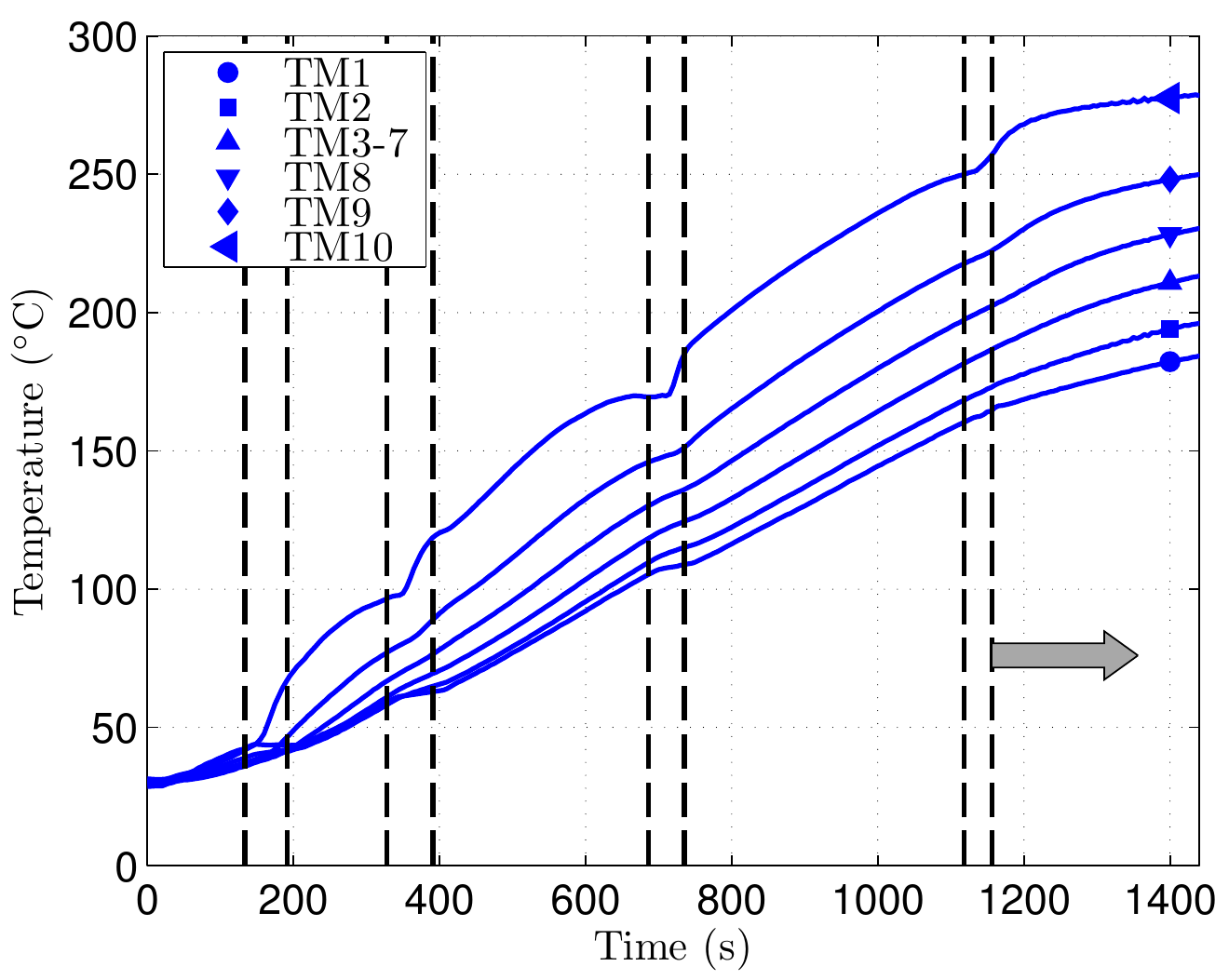}}
\subfloat[Workpiece temperatures \label{fig:TCBl}]{\includegraphics[width=0.5\linewidth]{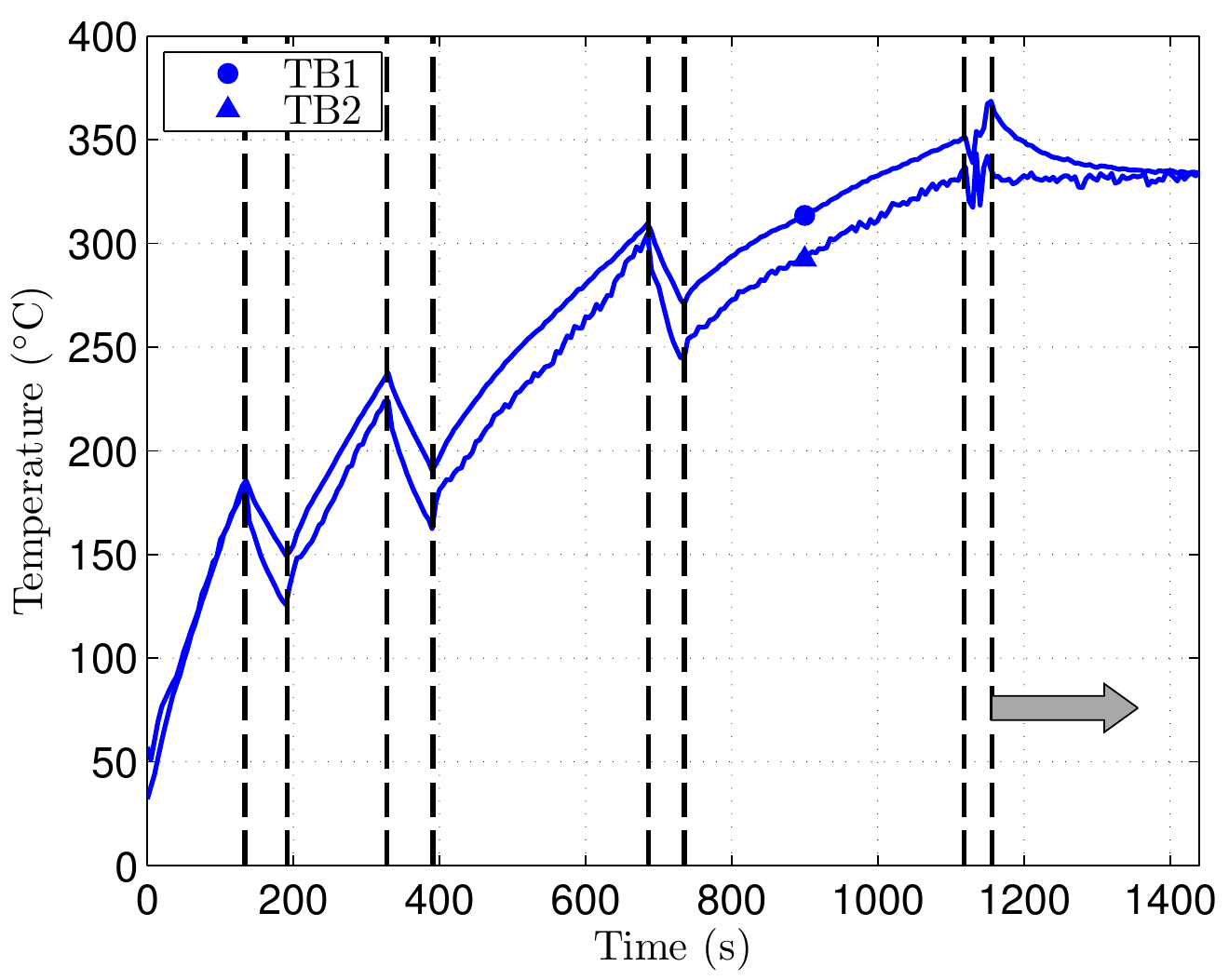}}
\caption[Mandrel and workpiece temperatures during preheating.]{Thermocouple record and axial location on both mandrel and workpiece employed for thermal characterization. Solid arrow in (b) and (c) indicates start of forming.}
\label{fig:ExpTemps}
\end{figure}

In order to characterize the thermal profile in the workpiece prior to forming, 2 type-K thermocouple (TC) junctions were peened into the surface of a workpiece, offset on the circumference by 30$^{\circ}$ at different axial locations (Fig. \ref{fig:TCloc}), and then the workpiece fitted to the mandrel. This workpiece was employed only to measure the characteristic thermal profile of the forming process, and was not subjected to forming. The TCs mounted on the workpiece were monitored in conjunction with the other 10 TCs embedded near the surface of the mandrel (Fig. \ref{fig:Mandrel} and \ref{fig:TCloc}) to measure the temperature variation with time in both the workpiece and mandrel during preheating and forming experiments.

Heat was applied from ambient conditions with the mandrel rotating at 20 RPM until thermal expansion caused the clamps to loosen, after which heating was interrupted, the clamps re-tightened, and heating resumed. Once the workpiece reached the forming temperature of $\sim$350$^\circ$C, the mandrel rotation rate was increased to the forming rate (281 RPM) to determine the effect of increased rotational speed on temperature.  The temperatures measured in both the mandrel and workpiece during this exercise are presented in Fig. \ref{fig:TCMand} and \ref{fig:TCBl}, respectively. The dashed lines indicate where the mandrel rotation and heating system were disengaged to tighten the clamps, and then heating resumed. The solid arrow indicates the point where the mandrel rotation rate was increased from 20 to 281 RPM.

This exercise showed that there was little temperature variation in the workpiece (within 20$^\circ$C), with the TC closest to the mandrel interface reading the coldest temperature. The channel exhibiting both the highest mean temperature and largest heating rate on the mandrel was located immediately below the workpiece, and the temperature and heating rate dropped axially. There was a 1-2$^\circ$C difference in temperature between the TCs offset circumferentially on the mandrel, and due to this minor difference, the temperature from these TCs has been presented as an average value (TM3-7 in Fig. \ref{fig:ExpTemps}). Increasing the mandrel speed from the preheating rate to the forming rate  decreased the heating rate in both the mandrel and workpiece. The mandrel temperature continued to increase, while the workpiece temperature remained nearly constant at the TB2 location over twice the length of time necessary for forming. The temperature difference between TB1 and TB2 decreased gradually. 

These observations have been used to define the initial and boundary conditions employed in the model, described subsequently.

\subsection{Forming methodology}\label{sec:FormingMethod}
Workpieces intended for forming did not have TCs installed owing to their interference with forming. Surface temperature measurements were performed manually on each workpiece at the TB1 and TB2 locations with a type-K TC surface probe (Omega model number 88108) every 3 minutes during heating, as well as immediately before and after forming. Workpieces were preheated to the target temperature of 355$\pm8^\circ$C using a bank of propane torches. Heating times were between 17-23 minutes. The variance in workpiece heating time is attributed to the mandrel fitment; workpieces better conforming to the mandrel surface took longer to heat up as heat transfer to the mandrel was improved in spite of the refractory-type coating. Once the workpiece was at the forming temperature, the mandrel speed was increased to 281 RPM and the roller was brought into contact with the workpiece at approximately 30 mm per minute. The longitudinal thread-cutting feed screw on the lathe was then engaged to move the roller axially at a rate of 0.21 mm per revolution (Fig. \ref{fig:formSpeedsFeeds}). Heating via the propane torches was continued during forming to lessen heat loss and maintain the workpiece temperature. These parameters were selected based on process information inferred from \citeauthor{Cheng.10} and the capability of the lathe.

Once forming was complete, the workpiece was removed from the mandrel and left to air cool, avoiding potential distortion that could result from quenching. The deformation in all experiments was such that the inner diameter of the deformed region on the workpiece did not contact the mandrel surface. During the course of the process, the workpiece temperature remained above 340$^\circ$C for all trials.

\subsection{Experimental results}
\begin{figure}
\centering
\begin{minipage}{0.5\linewidth}
\centering
\hfill \subfloat[Undeformed\label{fig:RawGeo_Undef}]{\includegraphics[width=0.8036\textwidth]{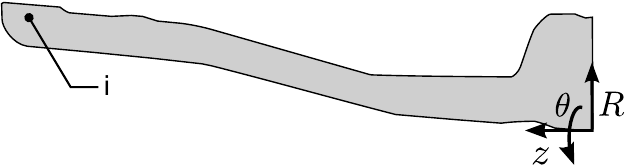}}\\
\hfill \subfloat[Low level deformation\label{fig:RawGeo_a}]{\includegraphics[width=0.8143\textwidth]{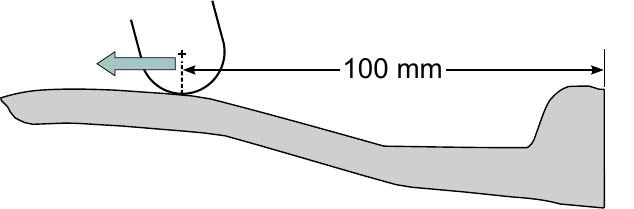}}\\
\end{minipage}\\
\begin{minipage}{0.4947\linewidth}
\centering
\hfill \subfloat[High level deformation\label{fig:RawGeo_b}]{\includegraphics[width=0.8929\textwidth]{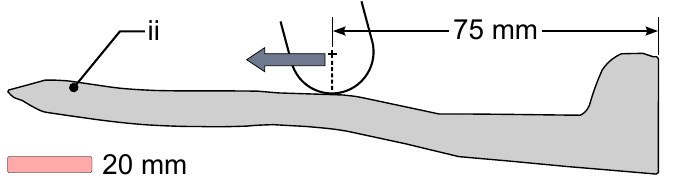}}\\
\end{minipage}
\caption{Cross-sectional geometries of an undeformed workpiece (a), low (b) and high (c) deformation workpieces. A depiction of the initial location and orientation of the roller nose is also provided on the cross-sections of the deformed specimens.}
\label{fig:RawGeo}
\end{figure}

In this paper, the results from forming two workpieces with increasing levels of deformation will be presented. The cross-sectional geometry of an undeformed workpiece as well as the two deformed workpieces were characterized by scanning the surface of a cross-section from each part and performing edge detection on the images. The profiles for each condition are presented in Fig. \ref{fig:RawGeo}. The formed specimen with the least deformation (Fig. \ref{fig:RawGeo_a}) corresponds to conditions normally seen in spinning operations, where plane strain effects dominate and little to no workpiece wall thickness changes were observed. The highly-deformed specimen showed an appreciable decrease in workpiece wall thickness (Fig. \ref{fig:RawGeo_b}), which is a departure from a standard spinning-imposed deformation.

\begin{figure}
\centering
\subfloat[As-cast]{\includegraphics[width=0.5\linewidth]{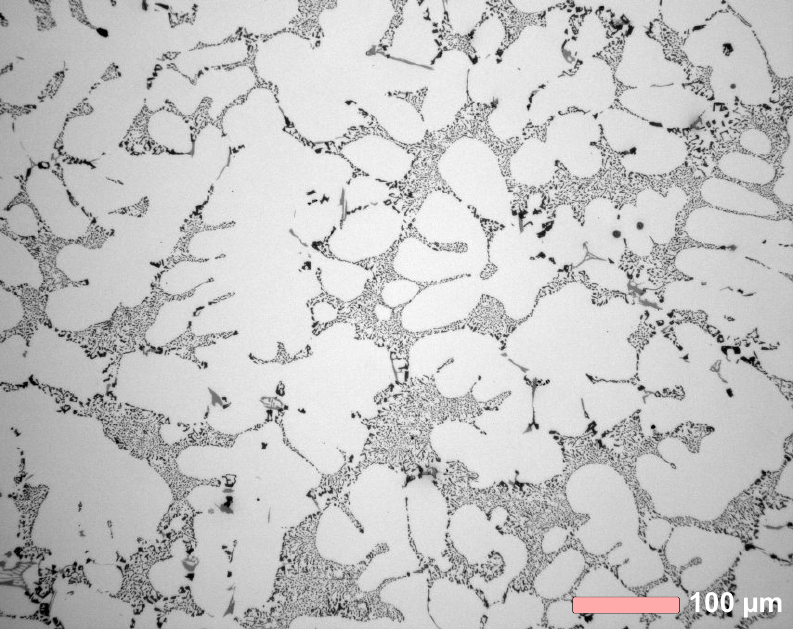}}\\
\subfloat[As-deformed]{\includegraphics[width=0.5\linewidth]{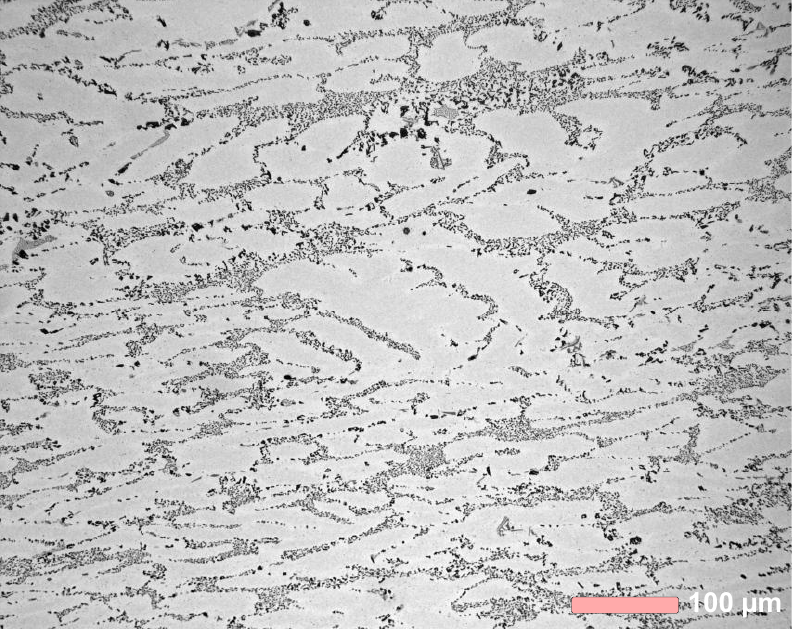}}
\caption{Initial as-cast microstructure in (a), and resulting microstructure in (b) after forming. Microstructure in (a) corresponds to position (i) in Fig. \ref{fig:RawGeo_Undef}, and microstructure in (b) corresponds to position (ii) in Fig. \ref{fig:RawGeo_b}.}
\label{fig:MicroStructure}
\end{figure}

The microstructure of the undeformed and deformed specimens was examined using optical microscopy once the surfaces of the cross-sections where polished. The deformed workpieces exhibited compacted dendritic structures, as demonstrated in Fig. \ref{fig:MicroStructure}. This change in microstructure is similar to that reported by \citeauthor{Mori.09}. It appears that grains with primary dendrite arms orthogonal to the forming direction have had the secondary dendrite arm spacing reduced as the grain is compacted. For grains with primary dendrite arms parallel to the forming direction, the secondary dendrite arm spacing has been increased as the grain conforms to forming.

\begin{figure}[]
  \begin{center}
  \subfloat[Crack locations]{\includegraphics[width=\linewidth]{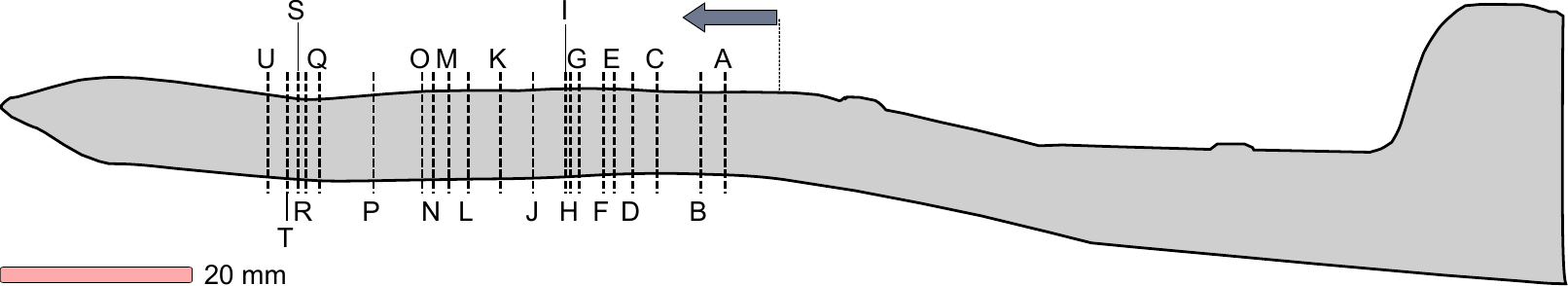}}\\
  \subfloat[Micrographs]{\includegraphics[width=\linewidth]{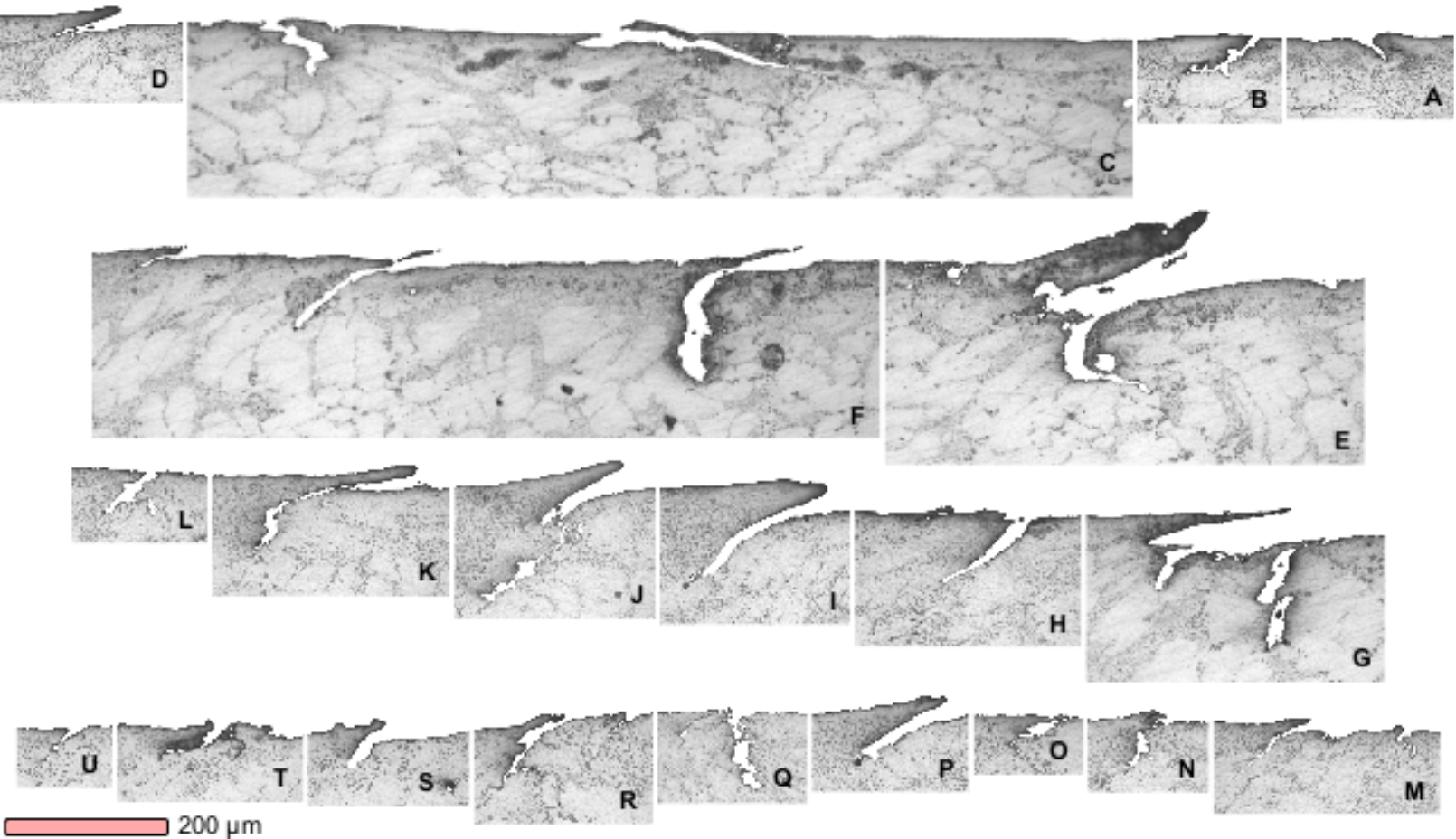}}
  \caption{Surface fracture details of heavily-deformed material. Micrographs are are aligned with the forming direction moving right to left.}
  \label{fig:MidCrack}
  \end{center}
\end{figure}

As demonstrated in Fig. \ref{fig:MidCrack}, the workpiece with the heaviest deformation showed surface defects in the form of cracking on the outer diameter in the worked regions. Cracking did not manifest on the low deformation sample. Cracks appeared on the surface of the sample early in the forming pass and alternated from opening counter to the forming direction (A, C) to predominately opening with the forming direction (B, D--U). The cracks do not extend any further than 140 $\upmu$m into the bulk of the material, as measured normal to the forming direction. With the exception of one crack (C) extending up to 320 $\upmu$m in length along the axis of the workpiece, the cracks tended to be short in axial length (not exceeding 200 $\upmu$m). The defects observed were larger in scale than those reported by \citeauthor{Mori.09}, who reported a maximum size of 60 $\upmu$m and these cracks were found to predominantly penetrate the surface of the sample with an angle of approximately 45$^{\circ}$ from the radial direction, occurring in eutectic-rich regions.

This type of localized failure is characteristic to the forming process, as it has been characterized in other forward rotary forming operations. Cracking or `fish scaling' in rotary forming is caused by highly localized shear occurring both ahead of and behind the roller interface. The macroscopic morphology of these types of defects are described by quite well by \cite{Rajan.01} for ferrous material. The extent of cracking is dependent on processing parameters and the local strength of the material. Both of these factors decide the overall crack morphology. Thus, the brittle eutectic phase appears to be a weak point in A356 where cracks initiate. This will be discussed subsequently in conjunction with the results of the model.

To eliminate cracking, either the forming temperature may be increased, or the forming parameters altered to avoid high levels of radial shear. Increasing the forming temperature from 350 to 400$^{\circ}$C was confirmed to arrest the formation of cracks by \citeauthor{Mori.09} for similar forming parameters. These results also agree with the study of A356 under torsion by \cite{McQueen.98}, who reported significant increases in strain to fracture moving from 300$^{\circ}$C to 400$^{\circ}$C, particularly at lower strain rates for this material. Thus, in addition to temperature, forming speed may be one of the process parameters that could be changed to reduce the frequency and severity of surface defects for rate sensitive materials.

\section{Coupled thermomechanical model development}
In order to calculate the evolution of stress, strain, strain-rate and temperature within the workpiece during rotary forming, a fully coupled thermomechanical model was developed using the commercial FEA software package, ABAQUS. The overall model included steps consistent with the experimental procedure outlined above by using submodels for workpiece preheating, forming at elevated temperature and cooling once forming was complete.

For the preheating model, axisymmetric heat transfer conditions were assumed to prevail in both the mandrel and the workpiece. Thermal and mechanical boundary conditions were imposed in order to calculate the preheated geometry and temperature distribution prior to forming. The results from this model were then translated to a 3D mesh of the workpiece. Forming operations applied to the workpiece were then modelled employing an adiabatic assumption using rigid analytical tooling to describe the mandrel and roller. The use of rigid analytical surfaces to describe the tooling was employed to reduce contact complications inherent with employing meshed instances of the tooling. Unfortunately, this assumption eliminates the ability to calculate the temperature history of the tooling. Mechanically, the choice of rigid surfaces is reasonable given the differences in flow stress and Young's modulus between A356 and tool steel at the forming temperatures. The results from the forming model were then used to populate a final submodel where the workpiece cooling in air was simulated to permit direct comparison between the experimentally achieved and simulated geometries at ambient temperatures.

Both implicit and explicit FEA methods were employed in simulating the overall process. Implicit analysis is best applied to problems without time-dependent geometric discontinuities as longer time steps can be used to converge on a particular solution. Meeting this criteria, the preheating model component used an implicit technique. However, for problems with large scale plasticity and involving contact, implicit methods are often unsuitable due to lack of convergence. In this latter case, explicit approaches are better suited as the solution is explicitly marched forward in time. As a result, the forming and cooling model components used explicit techniques.
\subsection{Geometry}\label{sec:Geometry}
The three main components considered in the model were the workpiece, mandrel and roller. During preheating, only the mandrel and workpiece were considered as 2D axisymmetric descriptions. Deformation conditions present during forming necessitated a complete 3D representation the stages occurring after preheating.

The rigid surface geometry used for the roller was sized according to the experimental description in Section \ref{sec:Exp} which was assumed to retain dimensions consistent with ambient temperature. This assumption was made because the roller was in contact only over a small area for a relatively short period of time, and any change in geometry due to the small temperature increase would be negligible. The geometry of the rigid surface used for the mandrel during forming was based on the temperature-corrected geometry of the mandrel calculated in the preheating submodel.

\begin{figure}
\centering
\begin{minipage}{0.5\linewidth}
\centering
\subfloat[Boundary conditions \label{fig:IC_BC}]{\includegraphics[width=\textwidth]{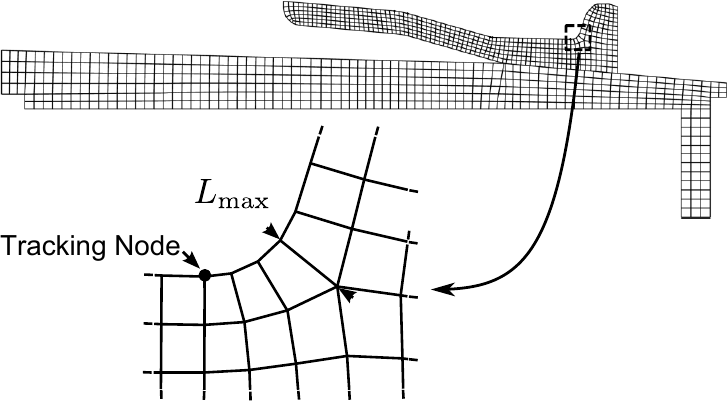}}\\ \hfill
\subfloat[Resulting deformed mesh \label{fig:IC_disp}]{\includegraphics[width=0.9336\textwidth]{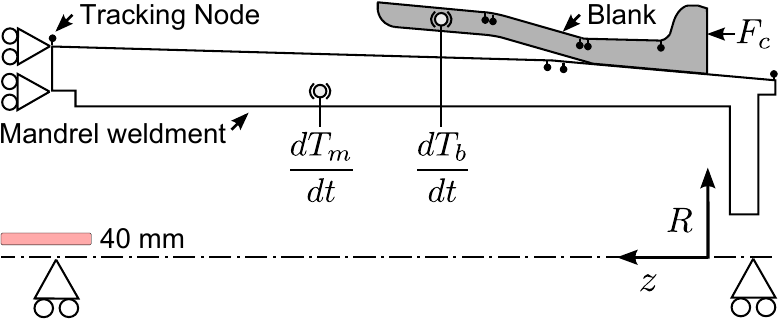}}
\end{minipage}
\caption{Preheating simulation boundary conditions and resulting deformed workpiece mesh used as initial conditions for forming.}
\label{fig:ICs}
\end{figure}

For the preheating submodel, a 2D section of the workpiece was decomposed into quadrilateral elements with a minimum element edge length of 2 mm. The resulting mesh had 573 nodes and 487 2D-axisymmetric, quadrilateral elements (Fig. \ref{fig:ICs}). The elements selected for the axisymmetric model featured hourglass control and reduced integration (CAX4RT). Five nodes on this 2D section were designated as tracking nodes to assist in later assigning boundary conditions for forming. A cross-section of the mandrel tooling was meshed with slightly coarser, elements having a minimum element length close to 4 mm, resulting in 673 nodes and 515 elements. 

Four tracking nodes were designated on the surface of the mandrel to track thermal expansion, in addition to the five on the outer surface of the workpiece. The simulated displacements of these nodes during preheating were used to define the rigid analytical surface of the mandrel and to define where the roller contacted the workpiece axially and radially.

To generate the mesh for use the in the forming submodel, the preheated 2D-axisymmetric mesh of the workpiece was revolved about its axis of symmetry to form a 3D mesh. In order to minimize the element count, the full 360$^{\circ}$ section of the workpiece was discretized into 297 circumferential segments, corresponding to the maximum element edge length of the 2D mesh ($L_{\max}\sim$  4 mm, Fig. \ref{fig:ICs}). This resulted in 170181 nodes and 144639 brick/hexagonal elements (C3D8RT), which inherited the reduced integration and hourglass controls from the 2D-axisymmetric elements. This mesh density was selected primarily on the basis of minimizing the computational resources while ensuring at least 4 elements were in contact with the roller surface during forming.

\subsection{Material properties}
The model required the elastic constitutive and thermophysical properties of both the mandrel and workpiece, as well as a description of the rate-dependent plastic behaviour of A356. Material properties were assumed to be isotropic.

An extended Ludwik-Hollomon constitutive expression developed by \cite{Roy.12} based on compression testing was implemented in the model via ABAQUS user subroutines. These subroutines calculate the local flow stress based on the temperature, strain and strain rate at each respective integration point for both the preheating (implicit) and forming (explicit) simulations. This material model was developed to describe equivalent stress based on experimental measurements at temperatures between 30 to 500$^{\circ}$C, strain rates between 0.1-10/s and plastic strain up to 0.35. The temperature-corrected elastic modulus and Poisson's ratio were incorporated as tabular data based on \cite{Frost.82}.

The thermal-physical properties of A356 were taken from \cite{Hetu.98}. Thermal conductivity and specific heat capacity ($k_c$ and $C_p$, respectively) as a function of temperature were implemented in the model using tabular data. The functions used to generate the data are shown in Table \ref{t:ThermProps}. A constant density, $\rho$, of 2670 kg m$^{-3}$ was assumed as the density of A356 does not change significantly over the range of processing temperatures. The expression for thermal expansion, $\alpha$, given by \citeauthor{Hetu.98} and provided in Table \ref{t:ThermProps}, was used to calculate the tangent thermal expansion data table as per the \cite{Abaqus.13}. 

In order to describe the thermal expansion of the mandrel solely during the preheating phase of the process, thermal-physical properties for AISI-4320 were necessary. The values for $k_c$, $C_p$, and $\alpha$ were entered in tabular form according to \cite{SteelProperties.90} and are summarized in Table \ref{t:ThermProps}. Density was taken to be 7850 kg m$^{-3}$ for AISI-4320, a modulus of 205 GPa and Poisson's ratio of 0.29 were prescribed.

\begin{table}
\caption[Thermal properties of A356 \& AISI--4320]{Thermal properties of A356 and from AISI--4320. Properties for A356 are those employed by \cite{Hetu.98} and AISI--4320 properties are from \cite{SteelProperties.90}.}
\centering
    \begin{tabular}{llll}
    \toprule
    & $C_p$ (J kg$^{\text{-1 }\circ}$C$^{-1}$) & $k_c$ (W m$^{\text{-1 }\circ}$C$^{-1}$) & $\alpha$ ($^{\circ}$C$^{-1}$)\\
    \midrule
    A356 & $898.7 + 0.4270T$ & $7146 + 4.150T$ & $2.260 \times 10^{-7}+\left(2.39 \times 10^{-8}\right)T$\\
    AISI--4320  & $452.1 +0.4740T$  & $43.33 - \left(1.090\times10^{-2}\right)T$ & $6.50\times10^{-7}$ \\
    \bottomrule
\end{tabular}
\label{t:ThermProps}
\end{table}

\subsection{Boundary and initial conditions}
Each of the simulations (preheat, forming, and cool-down) require mechanical and thermal boundary conditions to represent the sequence of steps taken experimentally, as previously described. Thermal boundary conditions were necessary to heat the workpiece and mandrel during preheating. During forming, the thermal boundary conditions on the workpiece were transformed to adiabatic conditions. Boundary conditions were then applied again to reflect cooling of the workpiece post-forming. Mechanical boundary conditions were used throughout these stages to control the relative movement of the workpiece and tooling and to describe the contact conditions.

\subsubsection{Preheating analysis}
In the preheating analysis, mechanical boundary conditions were applied in the form of node-based constraints on the spindle end of the mandrel section, such that axial movement was suppressed. This caused the mandrel to expand away from this constraint as the mandrel temperature increased. This constraint is consistent with the conditions present in the forming apparatus where the mandrel is rigidly fixed to end of the spindle and expands when heated towards the tailstock via the spring-loaded centre. A mechanical contact boundary condition was applied between the outer diameter of the mandrel and the inner diameter of the workpiece. Most rotary forming simulation studies have assumed frictionless contact, e.g. \cite{Mohebbi.10} and \cite{Xue.97}. A recent study conducted by \cite{Wang.11} on the unlubricated spinning of steel employed a Coulomb friction coefficient of 0.2 for tooling interactions. A Coulomb friction coefficient of 0.1 has been applied in the present work to all surface interactions as some friction is expected in spite of the graphite lubricant employed during experiments. A 10 kN force ($F_c$ in Fig. \ref{fig:IC_BC}) was evenly applied to the clamp face of the workpiece, causing the workpiece to maintain mechanical contact with the mandrel throughout the heating cycle and resulting thermal expansion. The load selected did not induce plastic deformation in the workpiece, but was sufficient to maintain contact.

Attempts were made to employ inverse thermal analysis to quantify the heat flux generated on the workpiece surface by the propane torches based on the experimental TC record (Fig. \ref{fig:ExpTemps}) and the domain described in Fig. \ref{fig:ICs}. It was found that the problem was severely ill-posed owing to the cropped mandrel domain, and the transient nature of the heat transfer between the workpiece and the mandrel. The variability observed in the heating rates between experiments further compounded difficulties with this analysis route. The heating rate variability was attributed to geometric variance in the workpieces, resulting in non-uniform axial and circumferential contact with the mandrel. As such, uniform heating conditions in the form of prescribed temperature boundary conditions were applied to the nodes corresponding to the mandrel and workpiece in the preheating model. These boundary conditions were prescribed to reach the measured temperatures of $T_m=220^{\circ}$C and  $T_b=350^{\circ}$C for the mandrel and workpiece, respectively, linearly ramping from an initial condition of 30$^{\circ}$C over 20 minutes. These boundary conditions ($dT_m/dt$ and $dT_b/dt$) are shown in Fig. \ref{fig:IC_BC}.

\subsubsection{Forming analysis}
\begin{figure}[]
  \centering
  \begin{minipage}{0.5\linewidth}
  \subfloat[Axial cross-section \label{fig:2DFormingModelBCs}]{\includegraphics[width=\textwidth]{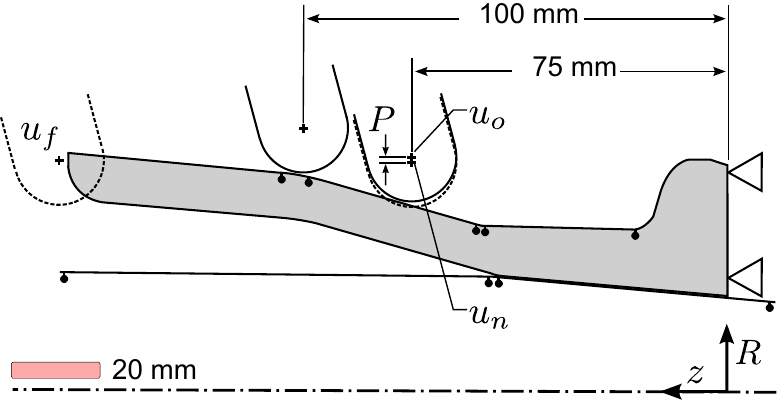}}\\
  \hfill
  \subfloat[Forming, 3D \label{fig:3DFormingModelBCs}]{\includegraphics[width=0.9571\textwidth]{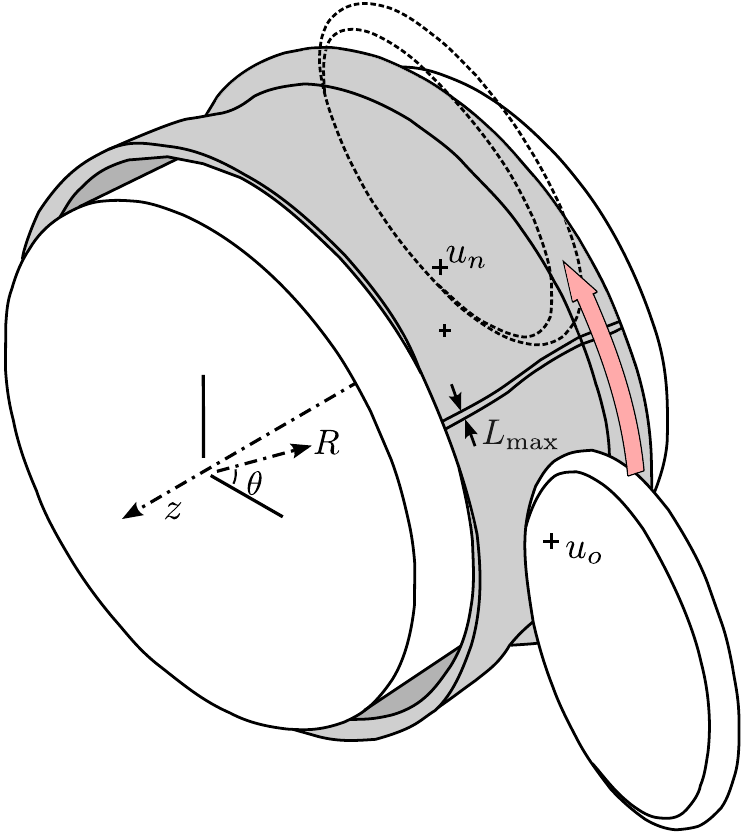}}\\
  \subfloat[Post-forming, axial cross-section \label{fig:CooldownBCs}]{\includegraphics[width=\linewidth]{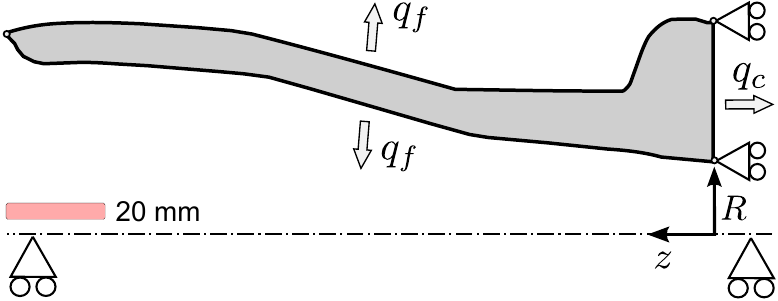}}
  \end{minipage}
  \caption{Forming model domain and boundary conditions: initial 2D axisymmetric model which was then translated to 3D for forming and post-forming simulations.}
  \label{fig:3DmodelDomainBCs}
\end{figure}

The thermal solution and geometry from the 2D preheating model was then translated to 3D (see Section \ref{sec:Geometry}) with the thermal solution employed as initial conditions for the 3D forming model. Mechanical boundary conditions for the workpiece were modified such that nodes on the clamp face were constrained, replacing the 10 kN force, $F_c$ (Fig. \ref{fig:IC_BC} versus Fig. \ref{fig:2DFormingModelBCs}).

Although the meshed instance of the mandrel was replaced by an analytical surface in the transition between preheating and forming, the mechanical contact boundary condition between the workpiece and the mandrel remained unchanged. An additional contact boundary condition was specified between the roller and the outer diameter of the workpiece. Adhering to the guidelines set out by \cite{Wong.08}, the workpiece was kept stationary and the roller was moved about and along its axis. 

During the forming experiments, the roller was moved into contact with the workpiece radially, and then set to move across the workpiece. In order to reflect this, the roller path in the forming simulation, starting with a small radial clearance from the outer diameter of the workpiece, initially moves radially towards the workpiece to make contact and then transitioned to a combined rotational and axial motion. The roller path was calculated based on the geometry and location of the preheated workpiece (on the mandrel) and then implemented in the model via a data table prescribing the motion of the reference node for the rigid analytical surface representing the roller.

In order to locate the roller path relative to the workpiece, the tracking nodes corresponding to the surface of the workpiece were used to define the geometry to match the two experimental forming profiles. The initial radial position of the roller nose was determined by linearly interpolating between the coordinates of the tracking nodes and an initial clearance of 0.1 mm. The initial axial position for each experimental condition was corrected for dilatation, and located on the geometry obtained from the preheating simulation. The path of the roller was then set to move first radially from the initial position $u_o$ to point $u_n$ with a penetration of $P=0.1$ mm, while rotating about the $z$-axis (refer to Fig. \ref{fig:2DFormingModelBCs}). Upon reaching $u_n$, the roller was then moved axially to point $u_f$ while rotating about the $z$-axis. The final axial position of the roller, $u_f$, was defined as the length of the corresponding experimental workpiece, corrected for temperature, plus 5 mm. This 5 mm clearance was imposed to ensure that the roller moved past the final point of contact with the workpiece at the end of the simulation. The proscribed axial movement of 0.21 mm per revolution and circumferential movement of 281 RPM of the roller matched those employed experimentally. This resulted in simulated forming process times of 61.93 and 87.76 s.

As mentioned previously, forming models were run adiabatically, to characterize the heat developed due to dissipation or inelastic heat generation without any heat loss from the workpiece. A Taylor-Quinney factor of $\beta=0.9$ was used.

\subsubsection{Cooling analysis}
Once the forming pass simulation was complete, the model was updated to reflect cooling conditions. Mechanical boundary conditions were imposed to prevent rigid body movement of the component during cooling, updating the conditions imposed during forming (Fig. \ref{fig:2DFormingModelBCs} versus Fig. \ref{fig:CooldownBCs}). The workpiece was then cooled down in the model by applying uniform surface heat fluxes to the surfaces of the workpiece (Fig. \ref{fig:CooldownBCs}). A flux of $q_f=4.3\times 10^3(T-25)$ W m$^{-2}$ was applied to both the inner and outer surfaces of the workpiece and $q_c=6.4\times 10^3(T-25)$ W m$^{-2}$ was applied to the clamped surface. With these fluxes applied, the workpiece cooled to ambient temperature in 7.5 seconds employing an explicit solution strategy. These fluxes were selected to limit the simulation time required to reach ambient temperature and to induce no further plasticity in the workpiece. Simulating the cooling step is necessary in order to compare the experimental workpiece dimensions at ambient temperatures with those predicted by the model.

\subsection{Computational overhead and scaling}
As described above, both implicit and explicit FEA methods were employed in simulating the process, with explicit forming and cooling model components requiring the highest computational overhead. This is because the maximum explicit time step is a function of the speed of sound through the shortest element edge employed.

A common approach to reduce computational resources for explicit analyses is to employ time or mass scaling. Both \cite{Mohebbi.10} as well as \cite{Wong.08} used these techniques to speed up solutions, but did not substantiate or justify the degree of scaling. This concept has been described in depth specifically for coupled FEA employing A356 by \cite{RoyTMS.12}. In order to determine the limits of scaling for the material model employed, Roy and Maijer used a select number of experimental thermomechanical tests which were employed to populate the material model and to validate a model of these tests using an implicit FE formulation. These implicit results and experimental data were used as points of comparison to explicit simulations with increasing levels of time and mass scaling. This study showed that time scaling of up to 75 times could be employed before incoherent flow stress between that modelled implicitly and experimentally obtained data was observed; all forming simulations in the current study employed a conservative 50x time scaling factor.

The shortest model computation times were arrived at by using parallel processing extensively on a modern multi-node computational cluster. While operating with a single processor, computation times were overly prohibitive; demanding at least 30 hours per second of scaled process time. The effect of additional processors significantly improved computation time, to a certain extent. Moving to two processors showed a 104\% improvement, but the rate of improvement showed diminishing returns with additional processors; moving from 7 to 8 processors only displayed a 2\% improvement with the additional processor. This is due to increased communication overhead as ABAQUS Explicit holds all element calculations on a single processor, and passes node-based calculations on to each additional processor. While adaptive meshing would be another method to improve solution times by maximizing element edge lengths and thereby maximizing time steps, ABAQUS does not support adaptive meshing on explicit coupled thermomechanical analyses.

The net result is that ABAQUS requires purely Langrangian 3D domains for fully coupled thermomechanical simulations, thereby demanding significant computational resources. Even with high levels of parallelization and without adaptive meshing, more than 3 hours of computation time were required per second of scaled process time. This corresponds to approximately 11 days for longest running model described above.

\section{Results and discussion}
Following the development of the model, it was run to infer the forming conditions achieved in the experiments. This model generated a large amount of information regarding the process including the ability to track the evolution of the stress state, strain rate, and temperature in the workpiece during forming. The results of the forming model reflecting the experimentally formed workpieces will be presented as an example of the model capabilities. These results will be followed with a comparison of the final workpiece geometries generated experimentally and those predicted with the model. Finally, the stress state imposed on the workpiece during high-level deformation will be presented and discussed within the context of surface defect formation.
\subsection{Model Results}
\begin{figure}[]
  \centering \setcounter{subfigure}{2} \centerline{
  \subfloat[End \label{fig:TopDownShort_end}]{\includegraphics[width=0.3259\linewidth]{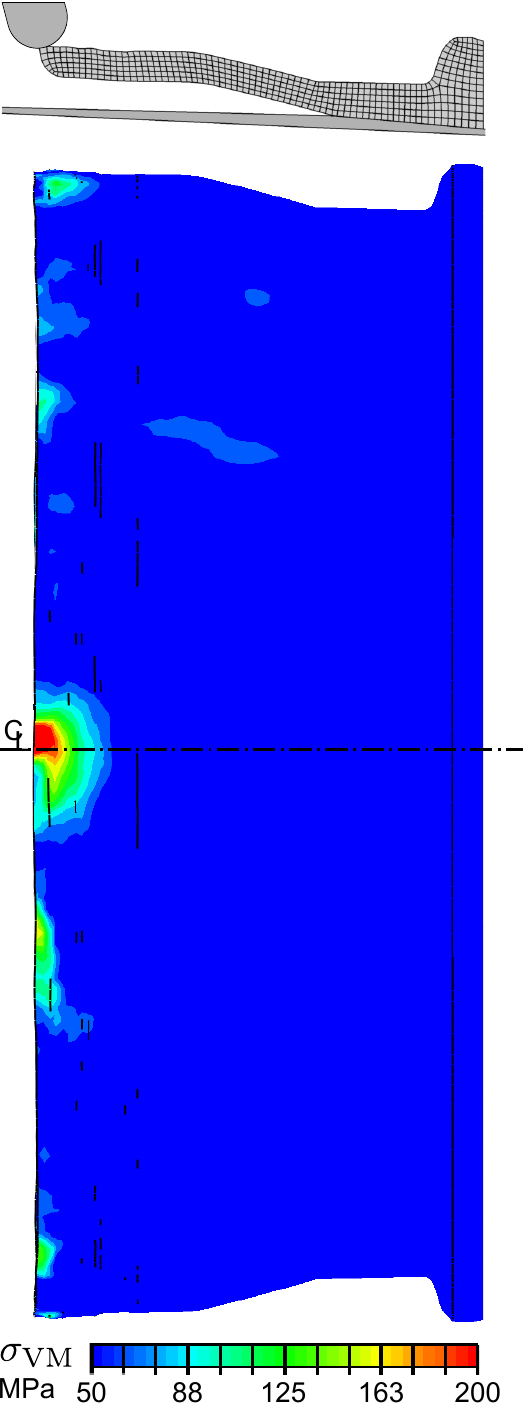}} \setcounter{subfigure}{1}%
  \subfloat[Midpoint \label{fig:TopDownShort_mid}]{\includegraphics[width=0.3259\linewidth]{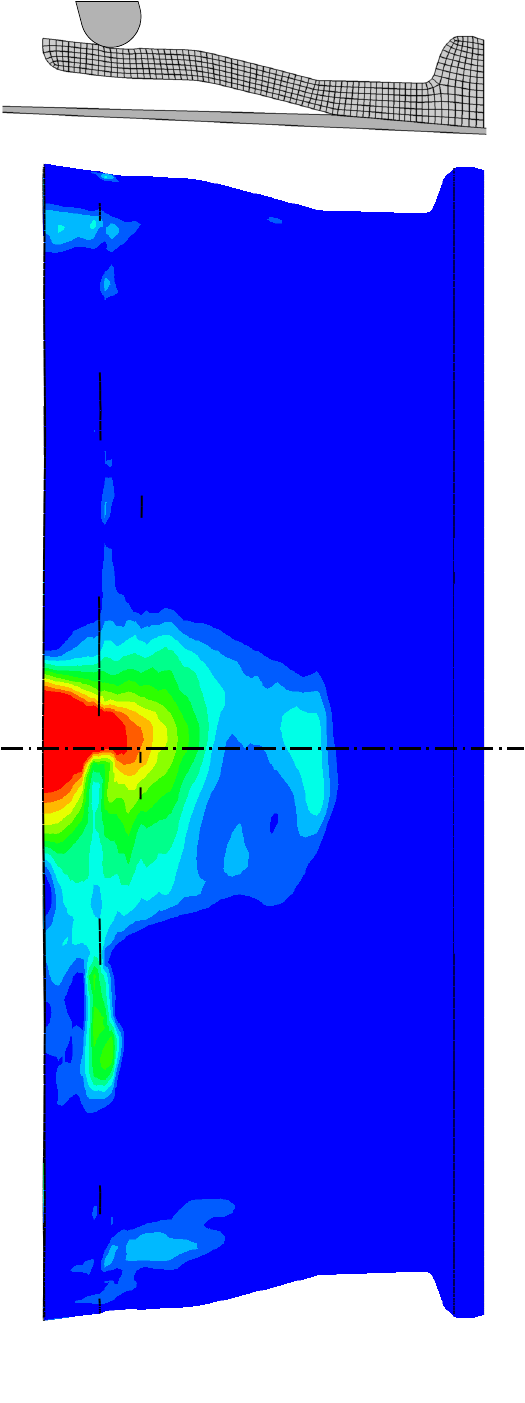}} \setcounter{subfigure}{0}%
  \subfloat[Start]{\includegraphics[width=0.3259\linewidth]{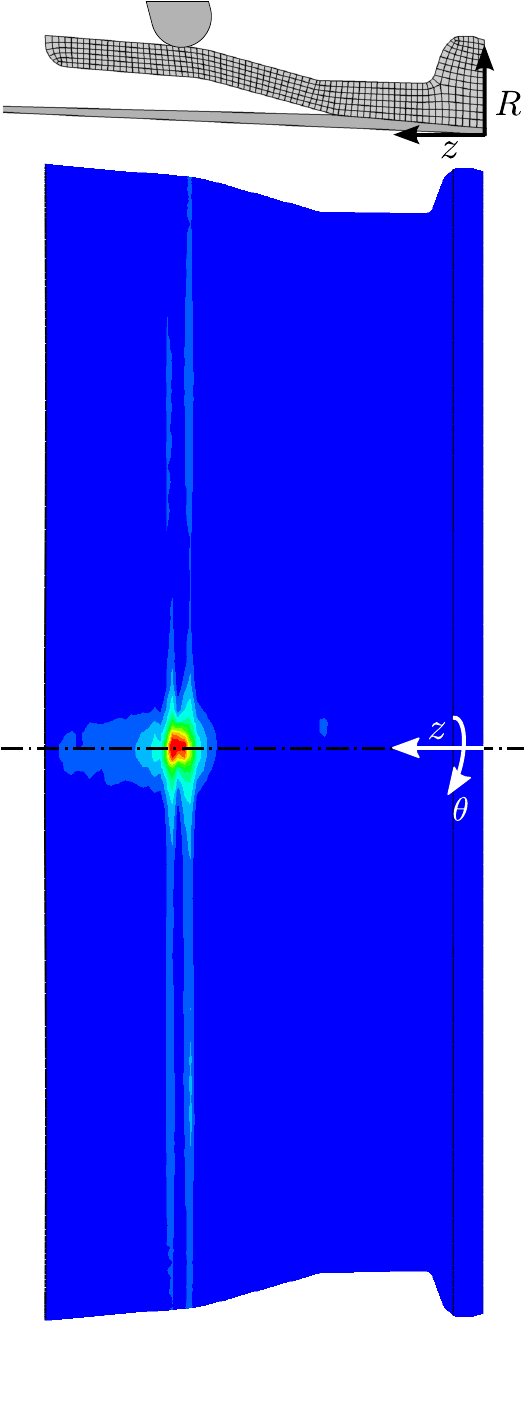}}}
  \caption{Predicted equivalent stress state ($\sigma_{\text{VM}}$) on the surface of the least-deformed workpiece during forming.}
  \label{fig:TopDown_short}
\end{figure}

The forming model provides a great deal of insight into the overall process. One facet is the evolution of the stress-state and deformation of the overall workpiece during forming. This is demonstrated in Fig. \ref{fig:TopDown_short}, which shows the distribution of the equivalent stress on the surface of the least-deformed workpiece at the start, midpoint and end of forming. This corresponds to three roller positions: i) $u_n$, ii) at the midpoint of the roller travel, and iii) just prior to the roller leaving the surface of the workpiece. In this figure, the workpiece has been rotated such that axis of symmetry of the roller is parallel with the centerline of the workpiece, affording an orthogonal view of the workpiece along its axis during forming. Here, and further, von Mises equivalent stress ($\sigma_{\text{VM}}$) has been used to analyze the workpiece stress state owing to the multiaxial nature of loading. 

Overall, the distribution of surface stresses are similar to those demonstrated by \cite{Xu.01} found by simulating the spinning of tubes: a very high compressive region in proximity to the roller acting radially, with a tensile region outside of this. However, due to the irregular shape of the workpiece employed, the forming zone is highly dynamic as compared to tube spinning. 

At the start of forming, the forming zone is highly localized and close to being symmetric about the centerline of the workpiece.  The majority of this zone, as identified by the stressed regions greater than 125 MPa, is directed circumferentially along the path of the roller, with the maximum contact stress appearing slightly behind the centerline, relative to the workpiece rotation. At the midpoint of forming, the stress state has evolved such that the majority of the forming zone remains ahead of the roller both axially and circumferentially, and significant stresses have evolved elsewhere on the surface. The region ahead of the roller and towards to the end of the workpiece along $z$ shows significant load directed axially. Also at this stage in forming, the overall bounds of the circumferential region carrying an elevated load has increased dramatically from the start of forming. At the end of forming, the stress state returns to being highly localized immediately beneath the roller, behind the centerline relative to the workpiece rotation.

\begin{figure}
\centering
\subfloat[Start]{\includegraphics[width=0.5\linewidth]{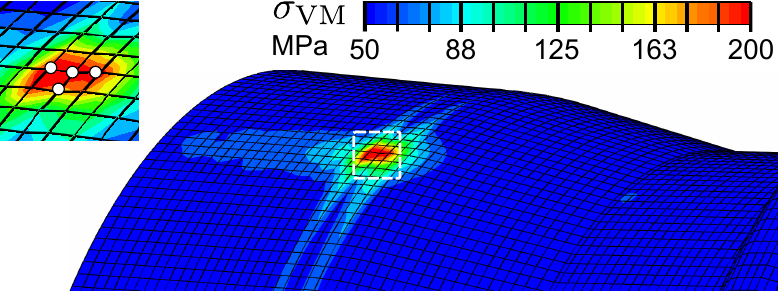}}\\
\subfloat[Midpoint]{\includegraphics[width=0.5\linewidth]{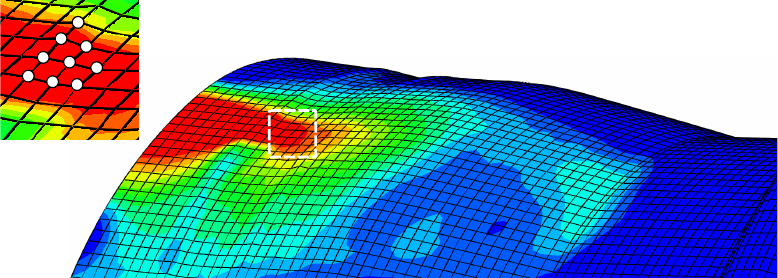}}\\
\subfloat[End \label{fig:ObliqueSDc}]{\includegraphics[width=0.5\linewidth]{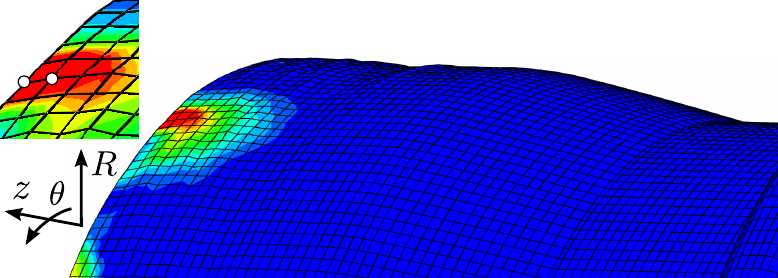}}\\
\caption{Oblique views of the simulated equivalent stress state immediate to the roller on the surface of the least-deformed workpiece. Inset shows nodes in contact with the roller.}
\label{fig:ObliqueSD}
\end{figure}

Fig. \ref{fig:ObliqueSD} shows an oblique view of the workpiece at the same stages during forming presented in Fig. \ref{fig:TopDown_short}. Also shown in this figure are the nodes in contact with the roller at each stage. At the start of forming, a faint ridge can be seen due to the initial contact of the roller, with a relatively small roller contact footprint. At the midpoint of forming, this ridge is more visible and contact has extended to include a small pileup of material ahead of the roller, which results in a larger effective contact region. The region ahead of the roller in the axial direction along $z$ demonstrates slight diametral growth, coinciding with the high axial stress levels as compared to behind the roller (along $z$). At the end of forming, the contact patch has diminished according with the dissipation of material pileup. In this last stage, the formed regions behind the roller exhibit ridges attributed to non-uniform pileup dissipation as the roller progressed across the surface of the component. 

\begin{figure}[]
  \centering \setcounter{subfigure}{2} \centerline{
  \subfloat[End \label{fig:TopDown_end}]{\includegraphics[width=0.3259\linewidth]{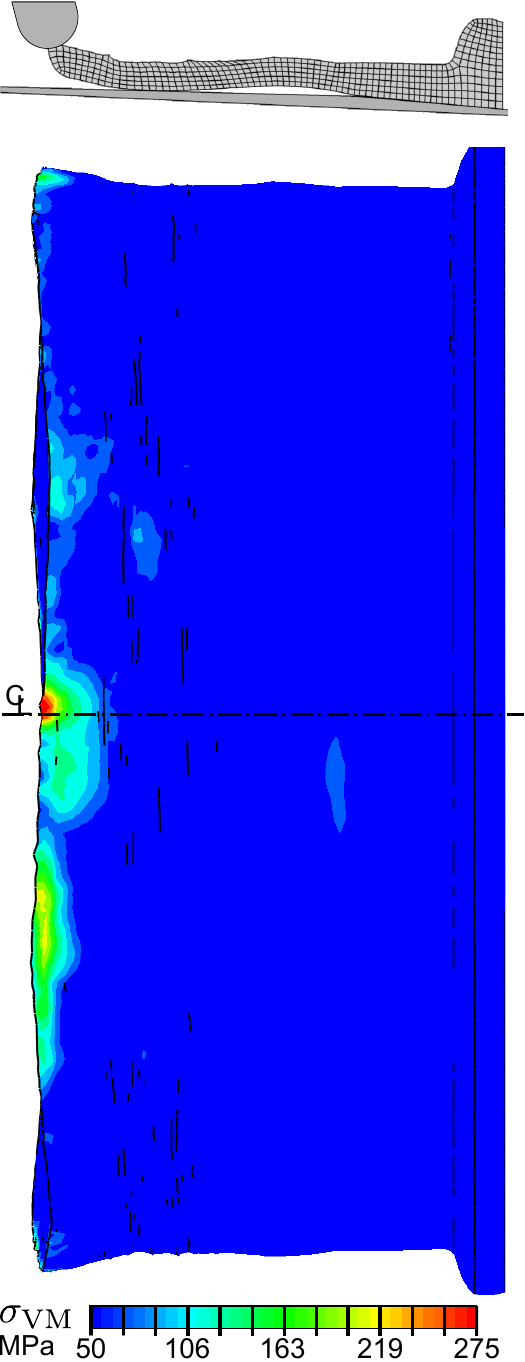}} \setcounter{subfigure}{1}%
  \subfloat[Midpoint \label{fig:TopDown_mid}]{\includegraphics[width=0.3259\linewidth]{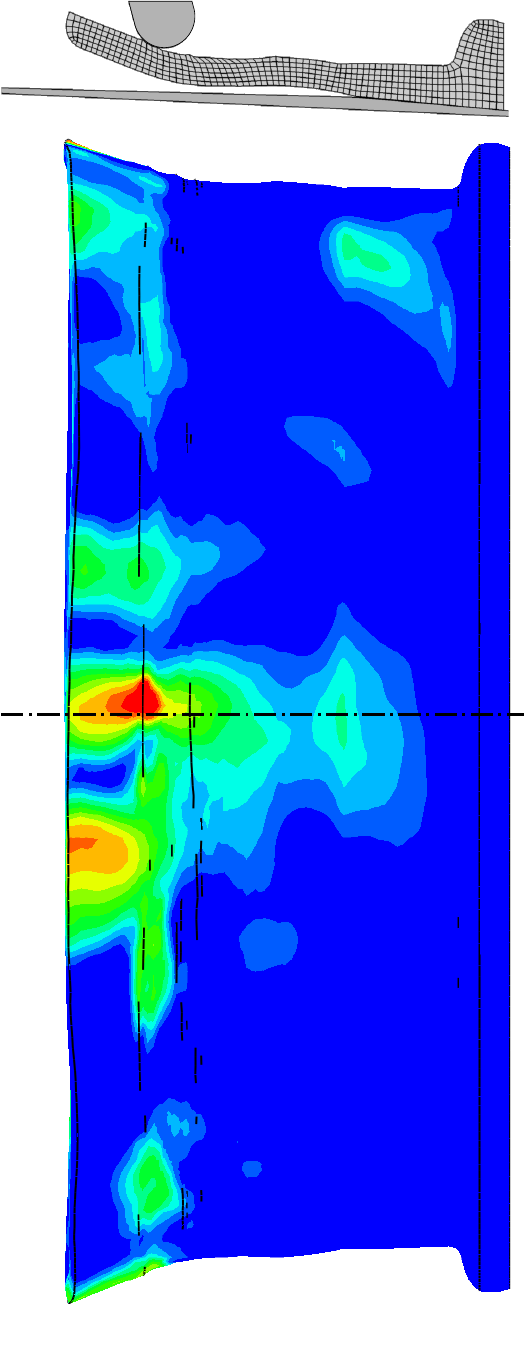}} \setcounter{subfigure}{0}%
  \subfloat[Start \label{fig:TopDown_start}]{\includegraphics[width=0.3345\linewidth]{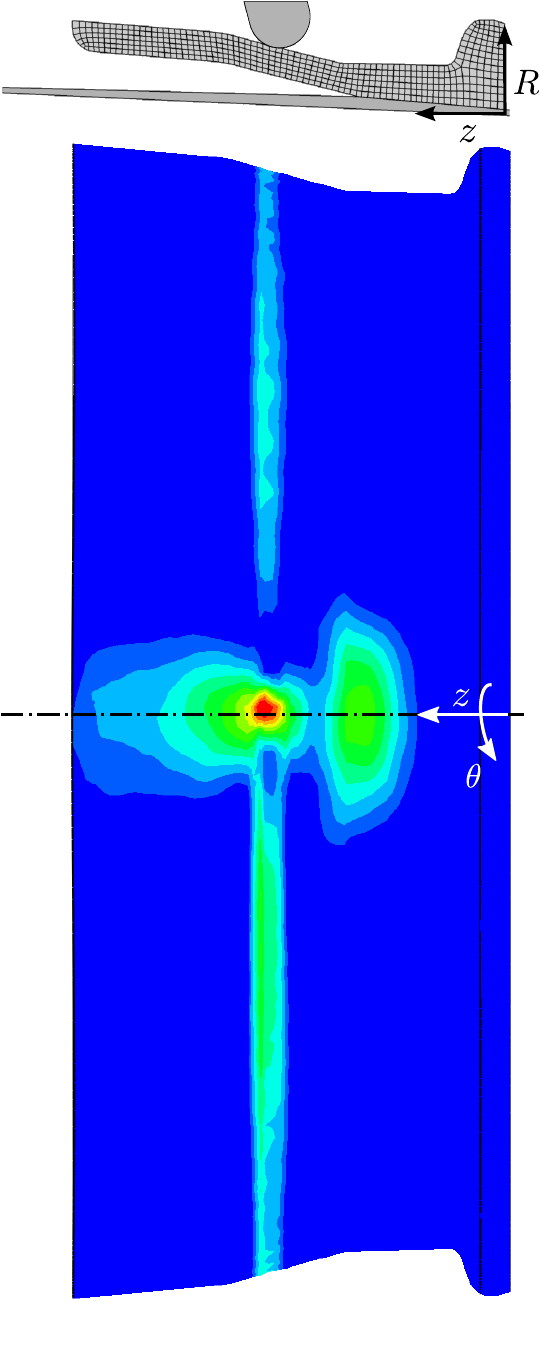}}}
  \caption{Simulated equivalent stress state ($\sigma_{\text{VM}}$) on the surface of the highly-deformed workpiece during forming.}
  \label{fig:TopDown}
\end{figure}

Orthogonal views of the deformation and stress states occurring in the highly deformed workpiece at the 3 stages during forming are presented in Fig. \ref{fig:TopDown}. Overall, the stress magnitude is significantly higher than seen with the least-deformed workpiece, coinciding with a more aggressive forming profile. Similar to the other workpiece, the peak stresses on the surface of the workpiece occur slightly ahead of the centerline at all forming stages. At the start of forming, the stress state is similar to the least-deformed workpiece in that the contact stress is approximately axially symmetric about the centerline of the roller. However, this stress state embodies a much larger region projected towards the fixed end of the workpiece, which is attributed to bending stresses caused by the roller contact. Midway through forming, the stress state has changed dramatically to be quite disparate from the least-deformed workpiece. This is attributed to the workpiece buckling and forming a convex flange ahead of the roller, which was observed experimentally. At this stage in forming, two regions of elevated stress appear on the workpiece circumference. One aligned with the centerline of the roller, and the other appearing on the flange, behind the roller circumferentially and ahead of the roller axially. At the end of forming, the stress state returns to being localized to the vicinity of the roller, however, the axial length shows a large degree of irregularity around the circumference, which was not seen in the least-deformed workpiece.

\begin{figure}
\centering
\subfloat[Start]{\includegraphics[width=0.5\linewidth]{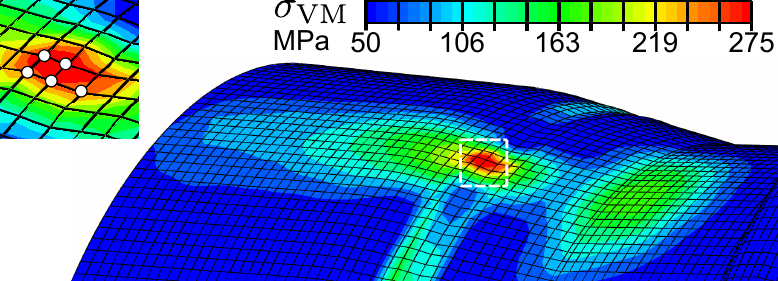}}\\
\subfloat[Midpoint]{\includegraphics[width=0.5\linewidth]{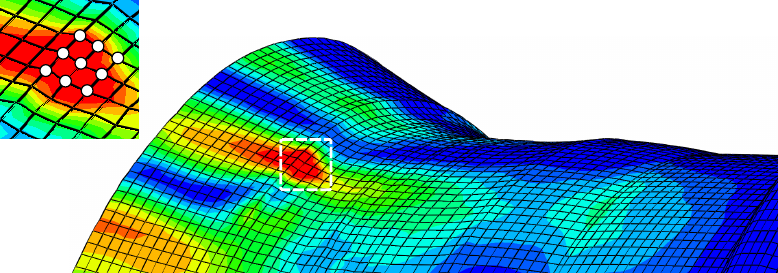}}\\
\subfloat[End \label{fig:ObliqueMDc}]{\includegraphics[width=0.5\linewidth]{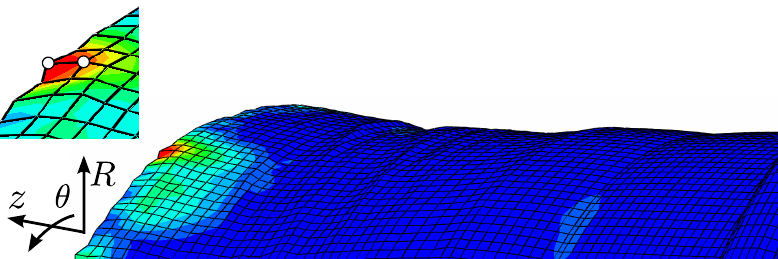}}\\
\caption{Oblique views of the simulated equivalent stress state immediate to the roller on the surface of the highly-deformed workpiece. Inset shows nodes in contact with the roller.}
\label{fig:ObliqueMD}
\end{figure}

Oblique views of the highly-deformed workpiece at the same forming stages are shown in Fig. \ref{fig:ObliqueMD} in conjunction with the nodes in contact. This shows that the same number of elements are in contact with the roller as the least-deformed workpiece at comparable forming stages. However, this is the sole similarity. The initial ridge formed by roller contact at the start of forming is much more pronounced. Beyond the formation of the flange midway through forming, the surface previously encountering the roller is much more irregular and a small region of pileup is seen at the edge of the flange closest to the roller. At the end of forming, the flange has collapsed, and the edge of the workpiece shows localized irregularities. The ridges seen in the least-deformed workpiece that were attributed to non-uniform pileup dissipation are much more exaggerated in this forming case. However, the ridges in the highly-deformed case are much less radially consistent than in the least-deformed workpiece. Clearly, the simulation predicts significantly less uniform deformation than observed in the least-deformed workpiece, with the workpiece wrinkling as the flange buckles.

\begin{figure}
\centering
\subfloat[Axial view]{\includegraphics[width=0.3397\linewidth]{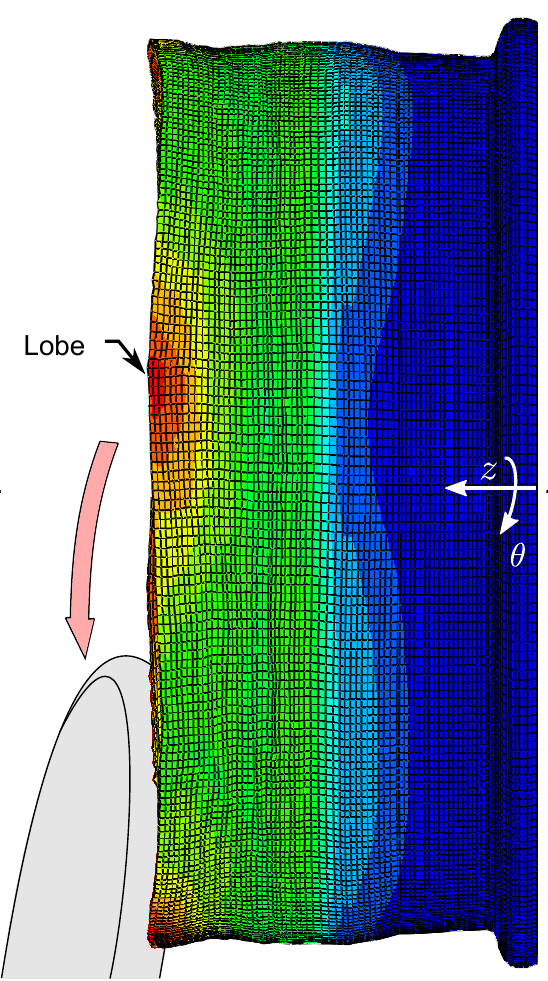}}\hfill
\subfloat[Normal to axis]{\includegraphics[width=0.6052\linewidth]{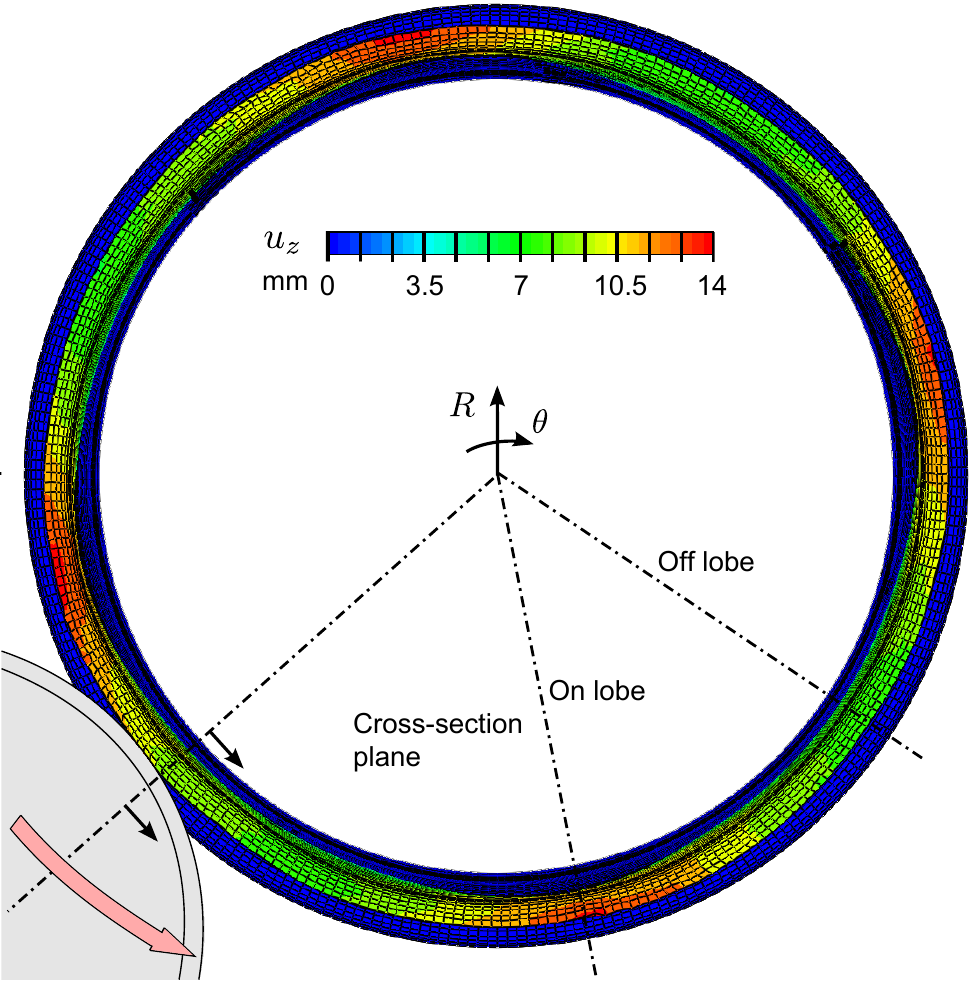}}
\caption{Contour plots of nodal displacement along the heavily-deformed workpiece axis, corresponding to Fig. \ref{fig:TopDown_end} and \ref{fig:ObliqueMDc}.}
\label{fig:3dMid}
\end{figure}

Two orthogonal views of the highly-deformed workpiece at the last stage of forming are provided in Fig. \ref{fig:3dMid} with contours showing the predicted axial displacement, $u_z$. This result highlights the non-uniform distribution of axial deformation, in particular the formation of lobes that are apparent at the end of forming. This phenomena is discussed further in the next section. To further examine the evolution of other process variables beyond stress and displacement, cross-sectional views of the simulated workpieces immediately underneath the roller have been extracted at each stage of the forming process. An example of one of these locations is shown in Fig. \ref{fig:3dMid}.

\begin{figure}[]
  \centering
  \begin{minipage}{0.49\linewidth}
  \subfloat[Start]{\includegraphics[width=\textwidth]{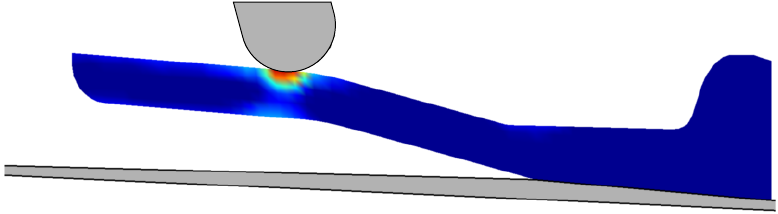}}\\
  \subfloat[Midpoint]{\includegraphics[width=\textwidth]{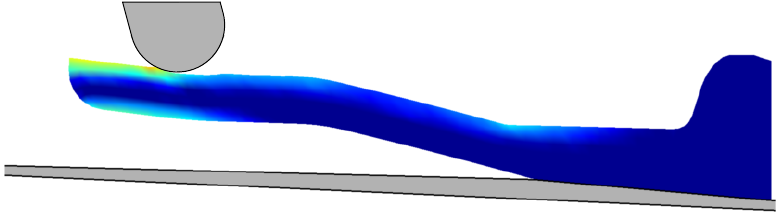}}\\
  \subfloat[End]{\includegraphics[width=\textwidth]{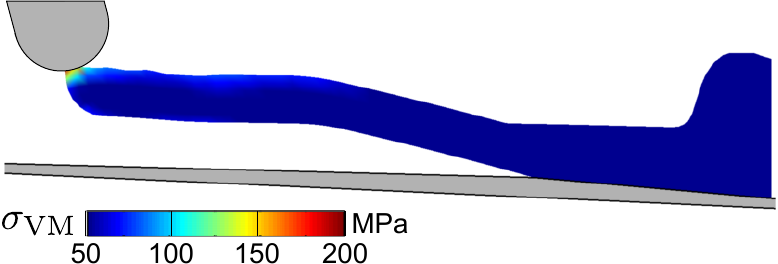}}
  \caption{Equivalent stress distribution on cross-sections of the least-deformed workpiece at different forming stages.}
  \label{fig:VM_axS}
  \end{minipage}\hfill%
    \begin{minipage}{0.49\linewidth}
  \subfloat[Start]{\includegraphics[width=\textwidth]{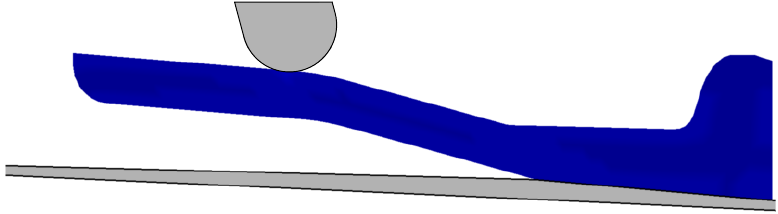}}\\
  \subfloat[Midpoint]{\includegraphics[width=\textwidth]{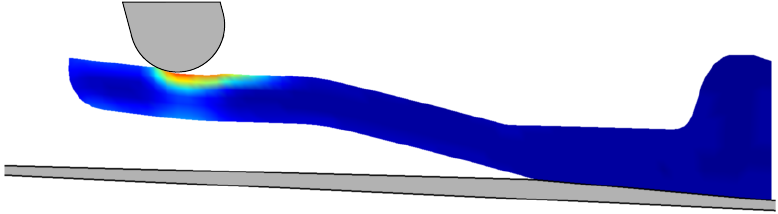}}\\
  \subfloat[End]{\includegraphics[width=\textwidth]{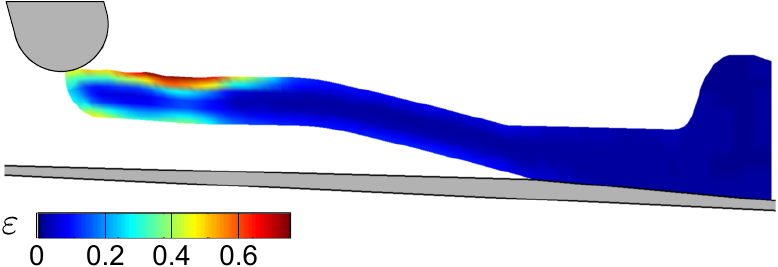}}
  \caption{Equivalent plastic strain distribution on cross-sections of the least-deformed workpiece at different forming stages.}
  \label{fig:eps_axS}
  \end{minipage}%
\end{figure}

\begin{figure}[]
  \centering
    \begin{minipage}{0.49\linewidth}
  \subfloat[Start]{\includegraphics[width=\textwidth]{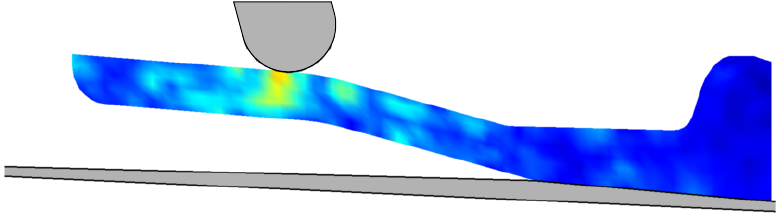}}\\
  \subfloat[Midpoint]{\includegraphics[width=\textwidth]{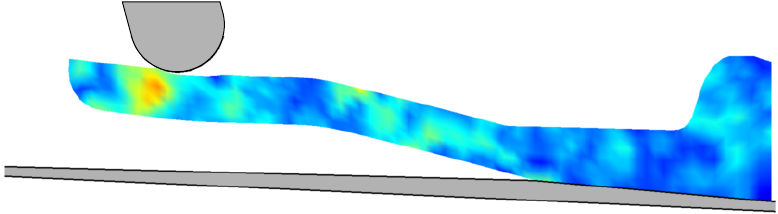}}\\
  \subfloat[End]{\includegraphics[width=\textwidth]{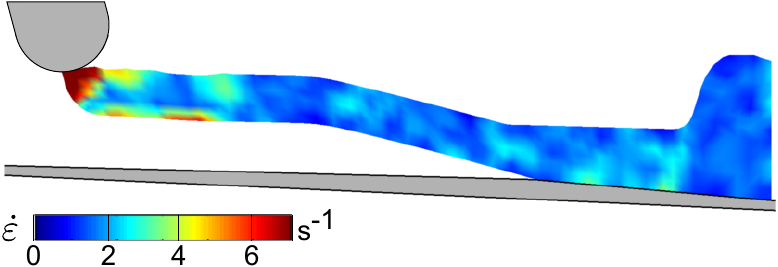}}
  \caption{Strain rate distribution on cross-sections of the least-deformed workpiece at different forming stages.}
  \label{fig:rate_axS}
  \end{minipage}\hfill%
  \begin{minipage}{0.49\linewidth}
  \subfloat[Start]{\includegraphics[width=\textwidth]{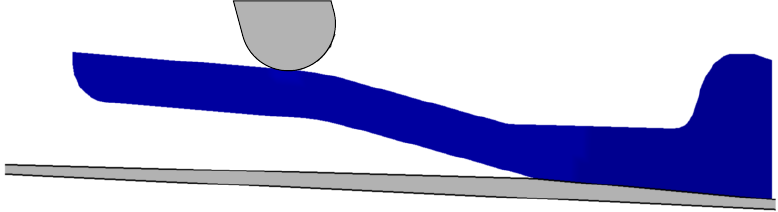}}\\
  \subfloat[Midpoint]{\includegraphics[width=\textwidth]{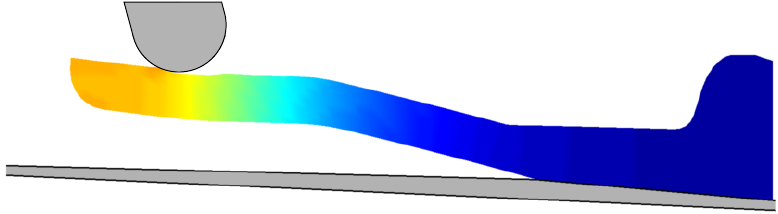}}\\
  \subfloat[End]{\includegraphics[width=\textwidth]{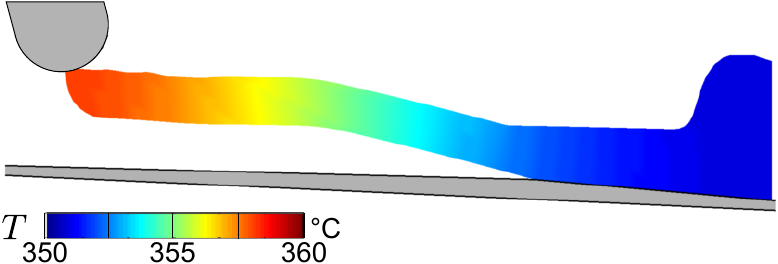}}
  \caption{Temperature distribution on cross-sections of the least-deformed workpiece at different forming stages.}
  \label{fig:T_axS}
  \end{minipage}%
\end{figure}

Contours of equivalent stress, equivalent plastic strain, strain rate, and temperature on the cross-sectional planes at each of the three stages during forming of the least-deformed workpiece are shown in Figs. \ref{fig:VM_axS} - \ref{fig:T_axS}, respectively. The equivalent stress in the least-deformed workpiece (Fig. \ref{fig:VM_axS}) at the start of forming shows that the region of elevated stress is localized directly beneath the roller and does not extend very far through-thickness. As forming progresses, the stress distribution evolves ahead of the roller, staying predominantly localized to the outer diameter of the workpiece as it bends to conform with the forming profile. The distribution of plastic strain (Fig. \ref{fig:eps_axS}) mirrors the stress profile, with a peak of 0.7 on the outer diameter and 0.3 on the inner. The strain rate distribution (Fig. \ref{fig:VM_axS}) which encompasses both elastic and plastic strain rates, has a peak located at the roller interface at the start of forming. Midway through forming, the peak strain rate has shifted to being slightly ahead of the roller and midway through the workpiece thickness. This is attributed to the combined effects of surface deformation and bending. The peak rate remains at approximately 4-4.5 s$^{-1}$ for most of the forming pass, rising to 7 s$^{-1}$ briefly at the end. Reflecting the relatively low amount of strain imparted to the workpiece, the temperature has only increased by approximately 10$^{\circ}$C (Fig. \ref{fig:T_axS}), less than 3\% of the initial temperature. This implies that for these forming conditions a small amount of heat was generated and a fully coupled model was not necessary (i.e. conditions were near isothermal).

\begin{figure}[]
  \centering
  \begin{minipage}{0.49\linewidth}
  \subfloat[Start\label{VM_mid_start}]{\includegraphics[width=\textwidth]{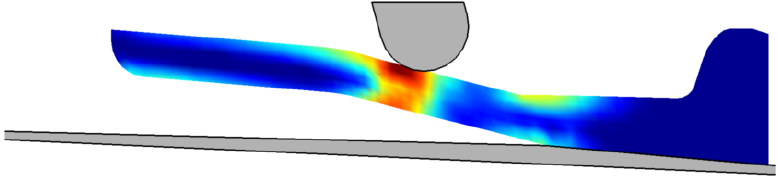}}\\
  \subfloat[Midpoint\label{VM_mid_mid}]{\includegraphics[width=\textwidth]{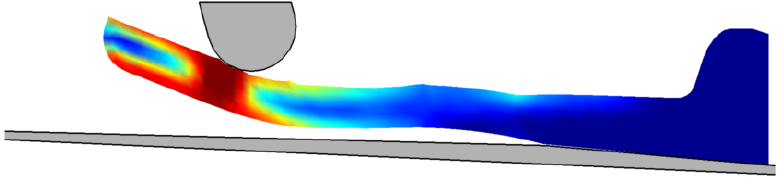}}\\
  \subfloat[End]{\includegraphics[width=\textwidth]{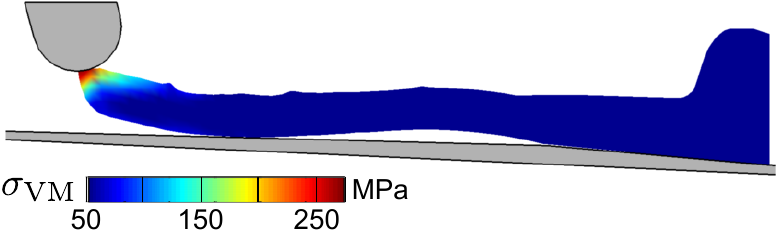}}
  \caption{Equivalent stress distribution on cross-sections of the highly-deformed workpiece at different forming stages.}
  \label{fig:VM_ax}
  \end{minipage}\hfill%
    \begin{minipage}{0.49\linewidth}
  \subfloat[Start\label{PE_mid_start}]{\includegraphics[width=\textwidth]{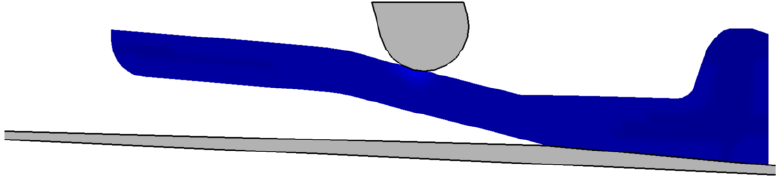}}\\
  \subfloat[Midpoint\label{PE_mid_mid}]{\includegraphics[width=\textwidth]{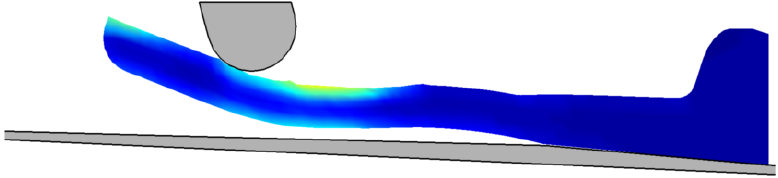}}\\
  \subfloat[End]{\includegraphics[width=\textwidth]{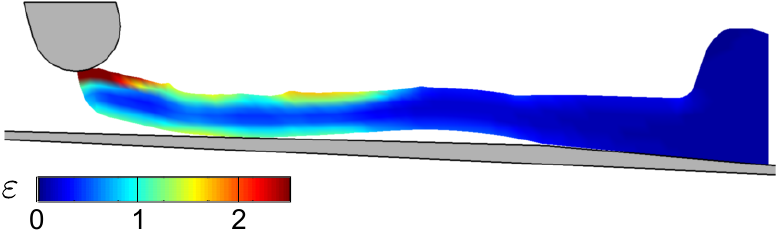}}
  \caption{Equivalent plastic strain distribution on cross-sections of the highly-deformed workpiece at different forming stages.}
  \label{fig:eps_ax}
  \end{minipage}%
\end{figure}

\begin{figure}[]
  \centering
  \begin{minipage}{0.49\linewidth}
  \subfloat[Start\label{ER_mid_start}]{\includegraphics[width=\textwidth]{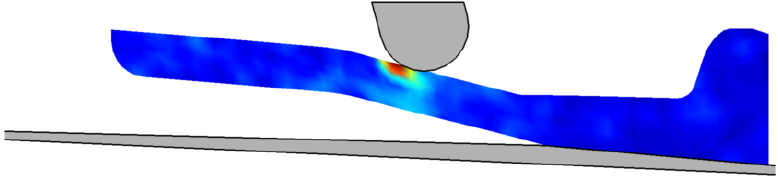}}\\
  \subfloat[Midpoint\label{ER_mid_mid}]{\includegraphics[width=\textwidth]{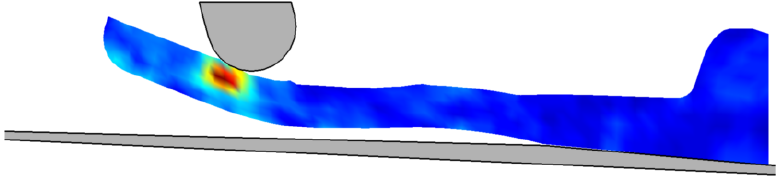}}\\
  \subfloat[End]{\includegraphics[width=\textwidth]{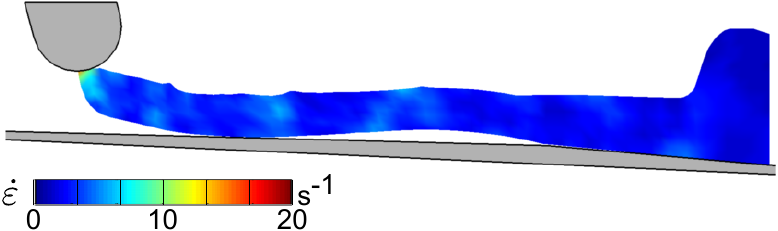}}
  \caption{Strain rate distribution on cross-sections of the highly-deformed workpiece at different forming stages.}
  \label{fig:rate_ax}
  \end{minipage}\hfill%
    \begin{minipage}{0.49\linewidth}
  \subfloat[Start\label{T_mid_start}]{\includegraphics[width=\textwidth]{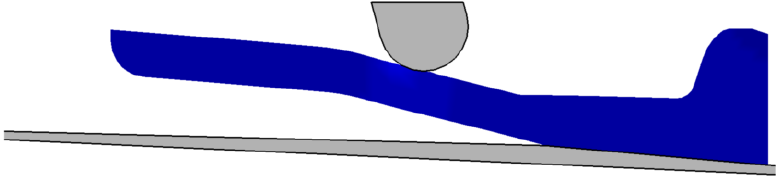}}\\
  \subfloat[Midpoint\label{T_mid_mid}]{\includegraphics[width=\textwidth]{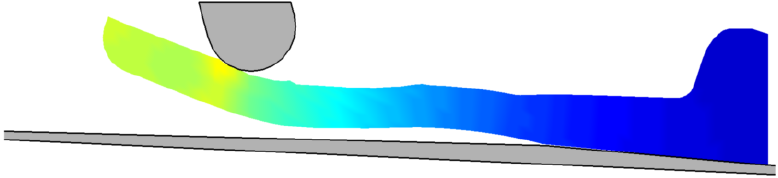}}\\
  \subfloat[End]{\includegraphics[width=\textwidth]{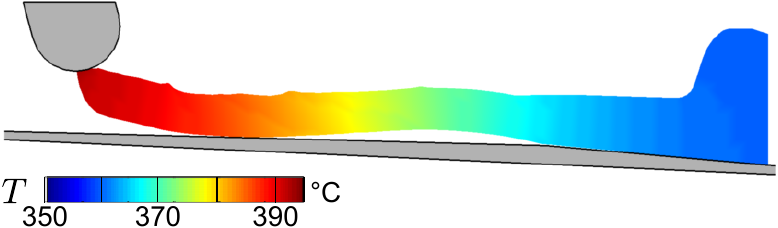}}
  \caption{Temperature distribution on cross-sections of the highly-deformed workpiece at different forming stages.}
  \label{fig:T_ax}
  \end{minipage}%
\end{figure}

Contours of equivalent stress, equivalent plastic strain, strain rate, and temperature on the cross-sectional planes at each of the three stages in forming of the heavily-deformed workpiece are shown in Figs. \ref{fig:VM_ax} - \ref{fig:T_ax}, respectively. The equivalent stress state occurring radially in the highly-deformed workpiece (Fig. \ref{fig:VM_ax}) demonstrates a much higher load both on the outer and inner diameter than the least-deformed workpiece. This is likely due to bending stresses developed as soon as the roller contacts the workpiece. As forming progresses, the stress state evolves ahead and to a lesser extent, behind the roller. The highest stresses appear in the middle of the workpiece, differing from the least-deformed workpiece as the deformation in the highly-deformed workpiece is dominated by bending. The distribution of equivalent plastic strain (Fig. \ref{fig:eps_ax}) is primarily localized on the outer and inner diameter of the workpiece at all forming stages. This is similar to the least-deformed workpiece, albeit peak strains are an order of magnitude higher. Additionally, at the midpoint of forming, an appreciable plastic zone has developed on the edge of the flange, well ahead of the roller, which is attributed to the start of buckling in the flange. At the end of forming, the flange region exhibits the largest amount of strain caused by the combination of buckling and roller contact during forming. The strain rate (Fig. \ref{fig:rate_ax}) distribution is similar to that observed in the least-deformed workpiece, however, the magnitude is approximately 4 times greater owing to the large amount of buckling. As there was significantly higher amounts of strain applied, the temperature increase is significantly higher, with the peak temperature increasing by 40$^{\circ}$C (Fig. \ref{fig:T_axS}). This rise in temperature has a significant effect on the flow stress of the material, and therefore a coupled model is necessary to accurately capture the material behaviour.

\subsection{Geometric Comparison to Experimental Results}
In order to gauge the effectiveness of the model to predict the final shape of the workpiece, the results of the model after cooling the deformed workpiece to room temperature have been compared in two ways. First, the predicted workpiece lengths are compared over the entirety of the circumference. Second, the predicted workpiece cross-sections are compared through thickness with corresponding experimental profiles (Fig. \ref{fig:RawGeo_a} and \ref{fig:RawGeo_b}).

\begin{figure}[]
  \centering
  \begin{minipage}[b]{0.49\linewidth}
  \includegraphics[width=\textwidth]{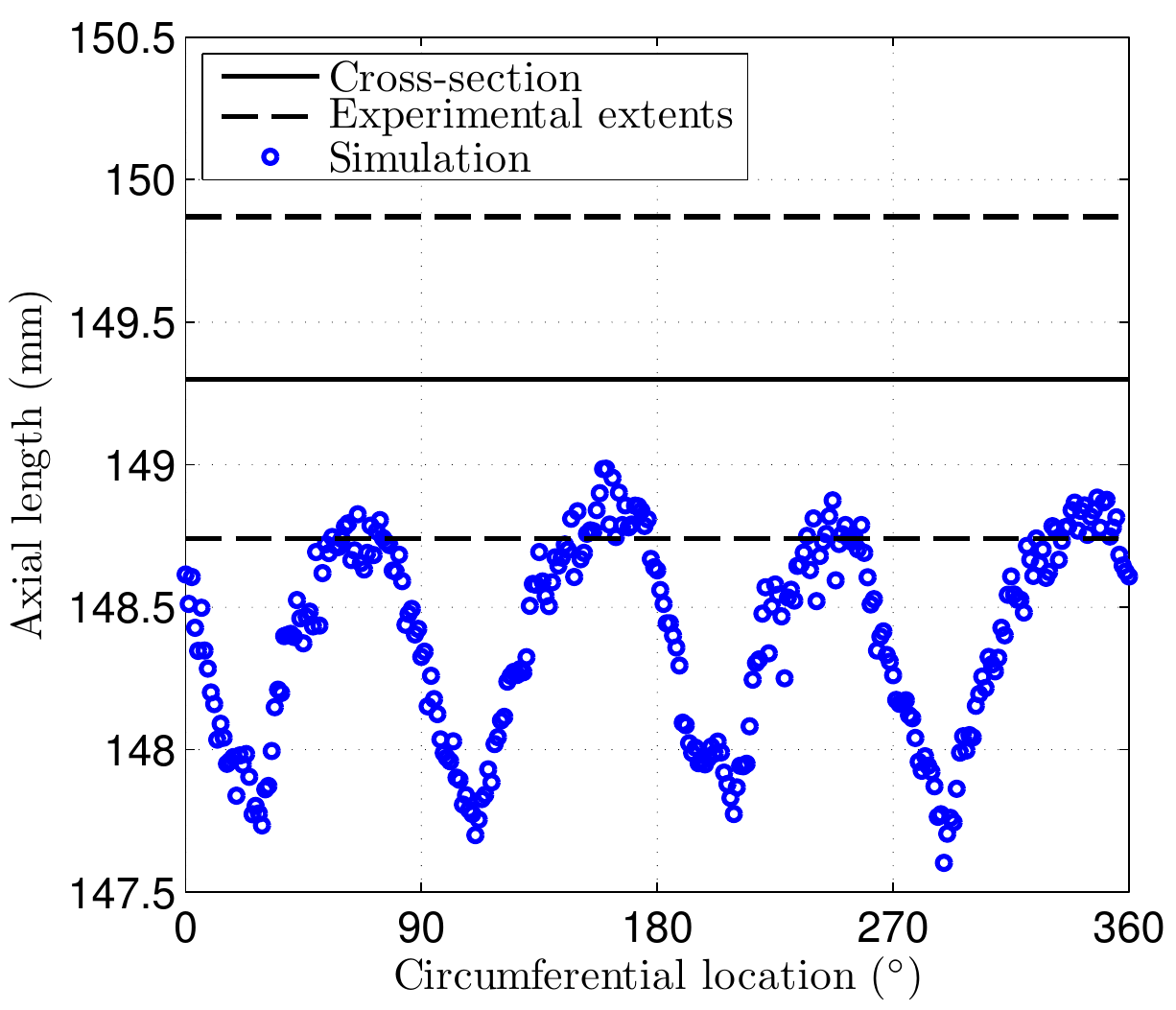}
  \captionof{figure}{Final axial length of the least deformed workpiece compared to experimental measurements.}
  \label{fig:axLengthCompS}
  \end{minipage}\hfill%
  \begin{minipage}[b]{0.49\linewidth}
  \subfloat[On lobe]{\includegraphics[width=\textwidth]{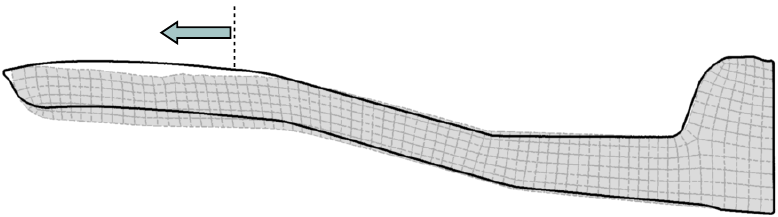}}\\
  \subfloat[Off lobe]{\includegraphics[width=\textwidth]{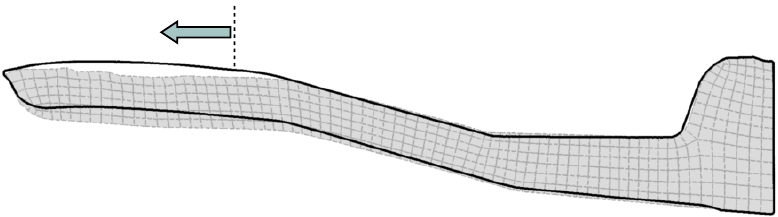}}
  \caption{Cross-sectional comparison of model and experimental results for least-deformed workpiece.}
  \label{fig:circumCompS}
  \end{minipage}
\end{figure}

Fig. \ref{fig:axLengthCompS} shows variation of the predicted axial length of the work piece around the circumference for the least-deformed case. The model data shows that there are four discrete minima and maxima appearing along the circumference of the part, corresponding to faint lobes. The measured extents of the least-deformed workpiece have been plotted as dashed lines for comparison. The solid line coincides with the overall length of the corresponding experimental cross-section (Fig. \ref{fig:RawGeo}). The experimental workpiece did not exhibit lobes as the minimum and maximum length were offset circumferentially by 180$^{\circ}$. However, with the experimental workpiece measuring 149.25$\pm0.55$ and the simulated 148.43$\pm0.62$, the agreement between the model and experiment is within the length of an undeformed element.

Fig. \ref{fig:circumCompS} compares the shape of the experimental cross-section (solid outline) with that predicted by the model (mesh) at the location of the maxima (on lobe) and minima (off lobe). The model predicts the development of a convex shape axially along the outer diameter, whereas the experimental profile is slightly concave. Both predicted profiles exhibit approximately the same amount of error in describing the experimental cross-section, with the inner radius being approximately 4.5 mm smaller at the midpoint of the deformed region than found experimentally.

\begin{figure}[]
  \centering
  \begin{minipage}[b]{0.49\linewidth}
  \includegraphics[width=\textwidth]{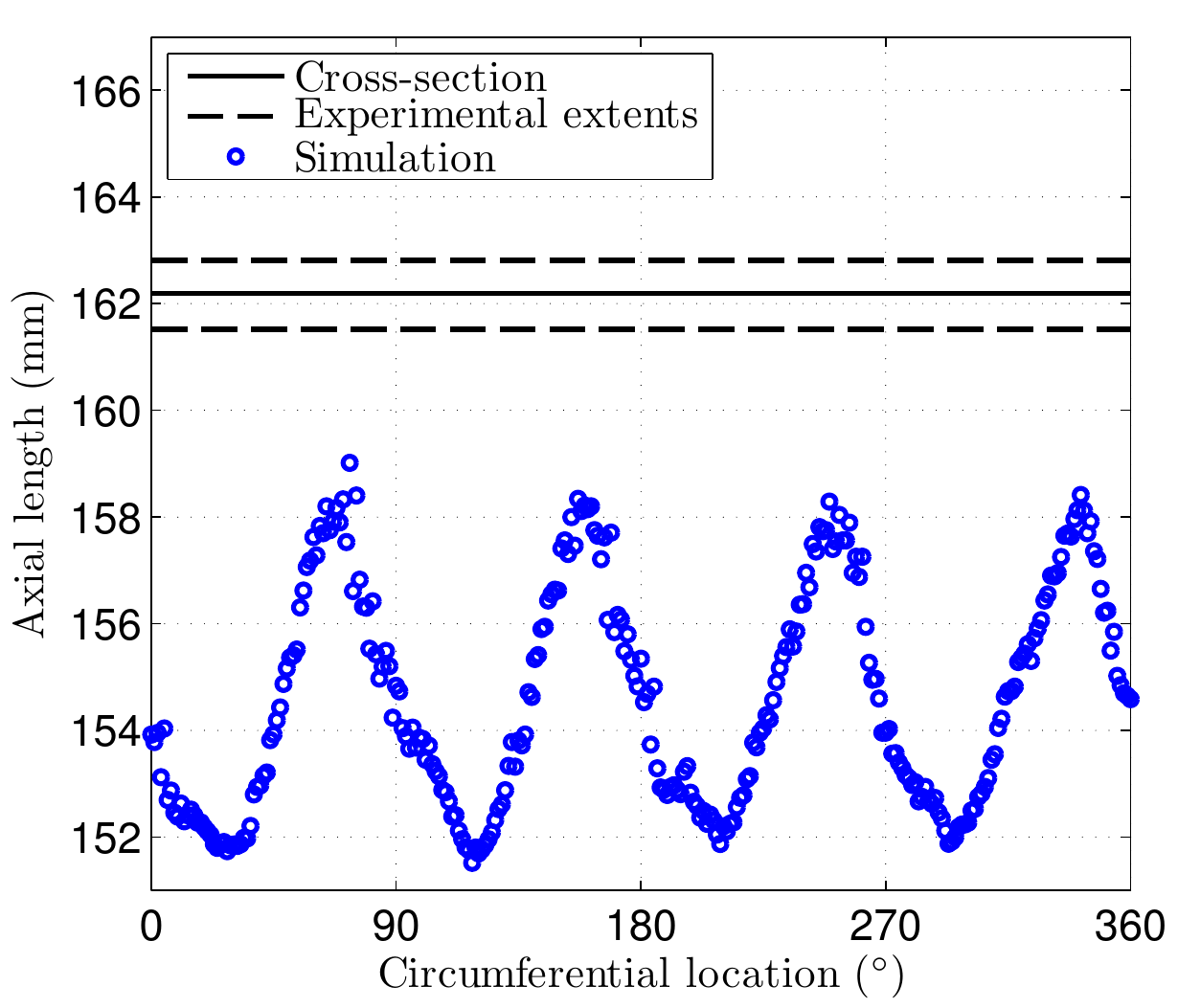}
  \captionof{figure}{Final axial length of the heavily-deformed workpiece compared to experimental measurements.}
  \label{fig:axLengthCompM}
  \end{minipage}\hfill%
  \begin{minipage}[b]{0.49\linewidth}
  \subfloat[On lobe]{\includegraphics[width=\textwidth]{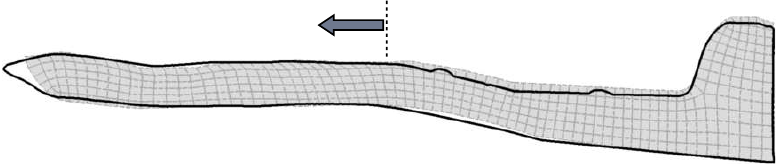}}\\
  \subfloat[Off lobe]{\includegraphics[width=\textwidth]{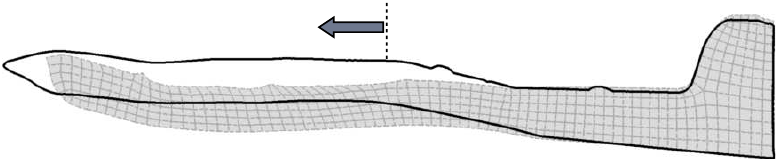}}
  \caption{Cross-sectional comparison of model and experimental results for the heavily-deformed workpiece.}
  \label{fig:circumComp}
  \end{minipage}
\end{figure}

Fig. \ref{fig:axLengthCompM} presents the same comparison as Fig. \ref{fig:axLengthCompS} for the heavily-deformed workpiece. For the high deformation conditions, the circumferential variation in length predicted by the model is much more pronounced, demonstrating much larger lobes. The final length of the experimental workpiece was 162.15$\pm0.65$ mm, where as the model predicts a length of 155.15$\pm3.15$ mm. The model does not predict the overall part length for the highly-deformed conditions nearly as well as the least deformed forming profile. This is also reflected in the comparison of the experimental cross-section with model predictions at the on and off lobe positions (Fig. \ref{fig:circumComp}). The on-lobe position agrees with the experimental cross section for the majority of the length, with the exception of the very end of the workpiece. The off-lobe position departs significantly, predicting an inner radius approximately 9 mm smaller at the midpoint of the deformed region.

The discrepancy between the model prediction and the experimental results for the heavily-deformed workpiece has been attributed to two potential sources of model inadequacy. First, there are limitations in the model to predict localized material failure. The model currently predicts localized deformation, but does not predict material fracture. Experimentally, surface cracks were observed on the outer diameter of the workpiece (Fig. \ref{fig:MidCrack}). Incorporating the effects of surface cracking in the model would affect the bending stresses significantly as the overall part stiffness would decrease. This would in turn modify the flange buckling phenomena. Based on the experimental results, it is surmised that surface cracking diminishes wrinkling and limits the formation of lobes. Second, the adiabatic conditions imposed may reflect an inaccurate account of how heat was transferred through the component during forming. This would induce a decreased yield point ahead of the roller, leading to inaccuracies in capturing the overall forming zone size, also modifying the flange buckling phenomena.

However, the model does show reasonable agreement with the experimental geometry in the case of the least deformed workpiece. It also provides an adequate prediction of the radial cross-section of the highly-deformed workpiece on lobes. The lack of agreement elsewhere for the highly-deformed workpiece can be attributed to cracking, and potentially inadequately imposed thermal boundary conditions. This suggests that the the basic deformation mechanism has been successfully modelled for some forming conditions.

\subsection{Surface defect formation}
As discussed previously, \citeauthor{Mori.09} demonstrated experimentally that surface cracks similar to those found in the present study occurred in regions of high levels of strain. These levels of strain were identified with a strain rate independent model. In reviewing the results of torsion testing conducted by \cite{McQueen.98}, fracture in A356 was observed to occur at strains of 1 and 1.5 at 300 and 400$^{\circ}$C, respectively, for strain rates up to 5 s$^{-1}$. The torsion tests also showed that the equivalent stress at fracture was seen to increase with higher strain rates.

Based on this information, the forming model can be used to explain the lack of obvious surface cracks appearing in the least deformed workpiece. The peak strain predicted by the model for this workpiece was 30\% less than the fracture strain identified by \citeauthor{McQueen.98}, at approximately the same temperature and strain rate conditions. In the case of the highly-deformed workpiece, the model predicts equivalent plastic strains of 1.5 or more during early forming, prior to the flange buckling. Furthermore, the predicted strain rate is significantly higher than the range employed by \citeauthor{McQueen.98}, which would decrease the fracture strain to a greater extent. However, strain and strain rate alone are not sufficient to predict local failure.

\begin{figure}[]
\centering
\includegraphics[width=0.5\linewidth]{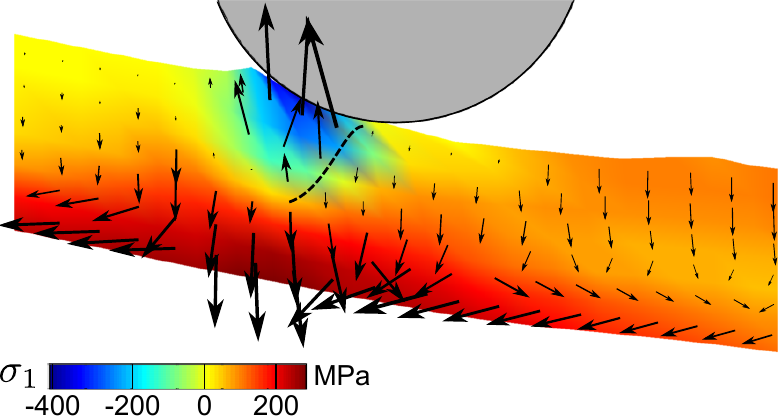}
\caption{Simulated $\sigma_1$ magnitude and orientation from the forming direction immediate to the roller. The depicted state was extracted from the simulation reflecting heavy deformation, at a quarter way through the axial travel of the roller, on the $z-R$ plane bisecting the roller.}
\label{fig:PrinStressJobbie}
\end{figure}

Fig. \ref{fig:PrinStressJobbie} shows a contour plot of the the principal stress ($\sigma_1$) magnitude occurring on a cross-section of the highly-deformed workpiece. Overlaid on the contour plot is a quiver plot showing the orientation of $\sigma_1$ relative to the workpiece axis, with origins located at the domain's integration points. This state reflects 17.5 mm of axial roller travel, or approximately 25\% through the forming profile, coinciding with a position halfway between those shown in Fig. \ref{fig:TopDown_start} and \ref{fig:TopDown_mid}. The region in contact with the roller demonstrates a high degree of compressive stress, however the element immediately behind the roller shows a nearly zero stress state at the surface. The stress state is increasingly tensile moving towards the inner diameter. In Fig. \ref{fig:MidCrack}, the crack morphology in terms of opening angle from the surface of the workpiece are quite similar. Notable exceptions include cracks A and C where the opening direction is reversed, but a much shallower crack is observed as compared to the majority. The most prevalent crack morphology matches the stress disparity as annotated by the black dashed line in Fig. \ref{fig:PrinStressJobbie}, reflecting unconstrained plastic flow.

Attributing the precise mechanical state shown in Fig. \ref{fig:PrinStressJobbie} to categorically identify the conditions necessary for crack development during forming is speculative at this stage in model development. This is because the material model employed is limited by the data available regarding fracture conditions. The model does, however, provide a framework to include further data with which forming parameters may be modified to conclusively mitigate this phenomena from occurring. One possible extension would be a criterion which tracks principal stresses to ascertain the degree of plastic constraint as a tool to predict whether or not cracking will occur. However, it will be necessary to experimentally develop the crack formation behaviour of this material at elevated temperatures.

\section{Summary}
In order to investigate spinning of full-size A356 components at elevated temperatures, an experimental forming apparatus was constructed. Two experimentally deformed components with different levels of spinning and buckling deformation were produced with this apparatus and were analyzed in this study. The workpiece with higher levels of deformation was found to exhibit surface defects in the form of cracking or `fish-scaling'.

The various thermal and mechanical processing steps involved in processing were modelled as a coupled thermomechanical process in ABAQUS. This consisted of an implicit FE submodel for preheating the workpiece to forming temperatures, as well as explicit FE submodels of the forming operation at elevated temperatures and final cooling to room temperature. The overall modelling effort contributes significantly to the overall understanding of rotary forming as a whole; heretofore, strain rate and temperature dependency in modelling efforts have not been considered in rotary forming, much less for A356.

The model shows that the local effects of the roller interface dominate throughout forming, and that the strain rates achieved are quite high, approaching that of standard metal turning. Changes in material state, manifesting with the evolution of strain rate and temperature produce non-uniform deformation in the form of wrinkling and the formation of lobes. Comparing the geometry of the lightly deformed workpiece with that predicted by the model shows reasonable agreement with the experimental geometry, demonstrating that the model successfully describes the basic deformation mechanism. Comparing the results for the highly-deformed workpiece showed that the model did not predict the geometry nearly as well. This may very well be due to the inability of the model to predict local damage. As additional data on cracking of A356 at high strain rates and strains at elevated temperatures becomes available, this model may provide the ability to modify forming parameters such that rotary forming defects may be avoided for this material.

As in a number of other previous studies which have attempted to model incremental rotary processes using commercial code, prohibitive computational requirements remain the primary limitation. This is particularly true when considering thermomechanical coupling. In order for FEA process modelling of rotary forming to be adopted by industrial practitioners for full-sized components, specialized code and/or custom hardware may be required. This is essential if modern techniques such as adaptive meshing are to be employed and to provide accurate results in a reasonable amount of time.

\end{document}